\documentclass[sigplan,10pt]{acmart}

\usepackage{times}
\usepackage{graphicx} 
\usepackage{subfigure} 

\usepackage{natbib}

\usepackage{hyperref}       
\usepackage{url}            
\usepackage{booktabs}       
\usepackage{amsmath,amsfonts}       
\usepackage{nicefrac}       
\usepackage{microtype}      
\usepackage{graphicx}
\usepackage{xspace,xcolor}
\usepackage{rotating}

\usepackage{tikz}
\usepackage{tikzscale}
\usetikzlibrary{positioning}
\usetikzlibrary{shapes.misc}
\usetikzlibrary{arrows.meta}
\usepackage{listings}
\usepackage{multirow}
\usepackage[linesnumbered, ruled]{algorithm2e}

\copyrightyear{2021}
\acmYear{2021}
\setcopyright{acmcopyright}\acmConference[PPoPP '21]{26th ACM SIGPLAN Symposium on Principles and Practice of Parallel Programming}{February 27-March 3, 2021}{Virtual Event, Republic of Korea}
\acmBooktitle{26th ACM SIGPLAN Symposium on Principles and Practice of Parallel Programming (PPoPP '21), February 27-March 3, 2021, Virtual Event, Republic of Korea}
\acmPrice{15.00}
\acmDOI{10.1145/3437801.3441578}
\acmISBN{978-1-4503-8294-6/21/02}

\setcopyright{acmcopyright}
\newcommand*{\origrightarrow}{}

\renewcommand*{\textrightarrow}{\fontfamily{cmr}\selectfont\origrightarrow}
\usepackage{booktabs}   

\usepackage{subfigure}
\begin{document}

\title[TurboTransformers]{TurboTransformers: An Efficient GPU Serving System For Transformer Models}       



\author{Jiarui Fang}
\affiliation{
\institution{Pattern Recognition Center, Wechat AI, Tencent Inc}            
\city{Beijing}
\country{China}                    
}
\email{jiaruifang@tencent.com}          

\author{Yang Yu}
\affiliation{
\institution{Pattern Recognition Center, Wechat AI, Tencent Inc}            
\city{Beijing}
\country{China}                    
}
\email{josephyu@tencent.com}          

\author{Chengduo Zhao}
\affiliation{
\institution{Pattern Recognition Center, Wechat AI, Tencent Inc}            
\city{Beijing}
\country{China}                    
}
\email{florianzhao@tencent.com}          

\author{Jie Zhou}
\affiliation{
\institution{Pattern Recognition Center, Wechat AI, Tencent Inc}            
\city{Beijing}
\country{China}                    
}
\email{withtomzhou@tencent.com}          

\renewcommand{\shortauthors}{Jiarui Fang and Yang Yu, et al.}


\begin{abstract} 
The transformer is the most critical algorithm innovation of the Nature Language Processing (NLP) field in recent years. Unlike the Recurrent Neural Network (RNN) models, 
Transformers can process on dimensions of sequence lengths in parallel, therefore leading to better accuracy on long sequences. 
However, efficient deployments of them for online services in data centers equipped with GPUs are not easy. 
First, more computation introduced by transformer structures makes it more challenging to meet the latency and throughput constraints of serving. 
Second, NLP tasks take in sentences of variable length.
The variability of input dimensions brings a severe problem to efficient memory management and serving optimization.

This paper designed a transformer serving system called TurboTransformers, which consists of a computing runtime and a serving framework to solve the above challenges.
Three innovative features make it stand out from other similar works.
An efficient parallel algorithm is proposed for GPU-based batch reduction operations, like Softmax and LayerNorm, major hot spots besides BLAS routines.
A memory allocation algorithm, which better balances the memory footprint and allocation/free efficiency, is designed for variable-length input situations.
A serving framework equipped with a new batch scheduler using dynamic programming achieves the optimal throughput on variable-length requests.
The system can achieve the state-of-the-art transformer model serving performance on GPU platforms and can be seamlessly integrated into your PyTorch code with a few lines of code.
\end{abstract}




\maketitle

\section{Introduction}
\label{sec:introduction}
The recent success of Nature Language Processing (NLP) techniques is enabled largely by the transformer-based Deep Neural Networks (DNNs), 
such as Seq2seq~\cite{vaswani2017attention}, BERT~\cite{devlin2018bert}, GPT2~\cite{radford2019language}, and XLNet~\cite{yang2019xlnet}, ALBERT~\cite{lan2019albert}.
With the support of the attention mechanism, the transformer models can capture long-range dependency in long sequences.
In data centers, GPU has proven to be the most effective hardware for deploying deep learning (DL) services.
The DL service accepts network requests and performs inference computation by performing a feed-forward pass on the model.
Although there are mature solutions for Convolutional Neural Network (CNN) and RNN services,
deploying transformer services with low latency and high throughput on GPU still faces two critical challenges.

Despite their success in model accuracy, transformer models are notorious for the massive amount of computation.
For the inference of a 40 words sequence, the base BERT model requires 6.9 Gflops.
To translate a 20 words sentence from Chinese to English, a typical Seq2seq model requires over 20 Gflops.
In comparison, for the inference of a 3x224x224 image, the ResNet50~\cite{he2016deep}, GoogleNet~\cite{szegedy2015going} and AlexNet~\cite{krizhevsky2012imagenet} require 3.9, 1.6 and 0.7 Gflops, respectively. 
Generally, transformer models lead to more computations than previous DNN models.

Besides enhanced computation requirement, transformer models introduced the problem of \textbf{variable-length} input,
where intermediate tensor dimensions have to change according to the input sequence length during a serving process.
Although also facing at variable-length input, RNN-based models, like LSTM ~\cite{hochreiter1997long} and GRU ~\cite{cho2014properties}, split the variable-length input into multiple fixed-length inputs and execute them sequentially.
Unlike models with fixed-length input, 
transformer models cannot benefit from pre-optimization of memory space allocated for intermediate tensors with known lengths, resulting in memory optimization challenges. 
In addition, due to variable-length input, the optimization of the serving framework becomes more complicated.
Conventional serving frameworks take advantage of batching techniques to increase GPU execution efficiency.
Extra computations brought by zero-padding of short requests in a batch of variable-length requests often conflict with the performance gains of batching computing.

Existing online serving solutions are not able to resolve both the large computation demand and the variable-length input issues, especially the latter one. 
The deficiencies can be summarized from the three following aspects. 
First, directly employing a training framework, such as TensorFlow ~\cite{abadi2016tensorflow} or PyTorch~\cite{paszke2019pytorch}, on inference tasks is not able to fully utilize the hardware resources. 
Inference differs from training in a way that it does not perform backward propagations which require extra memory space for intermediate tensors and eliminate the opportunity for operator fusion. 
Second, currently existing DL inference frameworks such as onnxruntime~\cite{onnxruntime}, 
TenorFlow XLA~\cite{xla}, TVM~\cite{chen2018tvm},  tensorRT~\cite{tensorrt} use techniques designed for fixed-length input workloads and have insufficient capability in dealing with variable-length input. 
Most of these current frameworks need a time-consuming preprocessing step to tune the computation pattern of the model operators according to their pre-determined input dimension. 
In order to work with transformer models, they usually convert variable-length requests into fixed-length requests through zero paddings, 
which not only introduces additional computational overhead but also limits the maximum length allowed to the pre-determined value. 
Although onnxruntime ~\cite{onnxruntime} recently provides some patches to support computation for variable-length input, 
the runtime's GPU memory management is not efficient.
After it serves a long request or a large batch of requests, a huge amount of memory allocated for intermediate tensors will not be released, introducing waste in terms of the memory footprint.
Third, none of the existing solutions have investigated serving optimization for variable-length input workloads. 
The request batching, which is the technique most helpful for performance, adopted in modern serving systems, such as TF-serving~\cite{tfserving}, Clipper~\cite{crankshaw2017clipper}, Nexus~\cite{shen2019nexus}, are suitable for only fixed-length input.

To solve the challenges of deploying an efficient transformer service, we proposed a serving system called TurboTransformers.
The system consists of a light-weight computation runtime and a serving framework.
The runtime adopts a variable-length-input-friendly design to avoid time-consuming preprocessing specified with dimensions of the intermediate tensor.
After loading a pre-trained model, the runtime rewrites the computation graph by fusing non-GEMM kernels,
and provides efficient CUDA implementations for them.
Before launching an inference, it conducts light-weight memory usage optimizations according to the input sequence length.
The runtime can achieve state-of-the-art speed and a smaller memory footprint compared with existing DL runtimes.
Moreover, it is easy to use, It can bring end-to-end speedup by adding a few lines of Python code.
Building upon the runtime, the serving framework improves throughput of the service through a variable-length-aware batching technique.

The innovative contributions of the paper are listed as follows.

\begin{itemize}
\item
We proposed a new parallel algorithm for Batch Reduction kernels like Softmax and LayerNorm,
which pushes the efficiency boundary of these kernels on GPU.
\item
We proposed a sequence-length-aware memory allocation algorithm.
Unlike other popular allocators for DNN runtimes, our allocator can utilize the computation-graph of the DNN model to derive efficient memory reusing strategies for variable dimension intermediate tensors.
\item
We proposed a sequence-length-aware batch scheduler.
It utilizes a dynamic programming algorithm to derive a batching scheme to achieve the optimal throughput.
\end{itemize}

\section{Backgrounds}
\label{sec:background}
\subsection{Transformer Models}
\begin{figure}[h]
\centering
\includegraphics[width=0.4\textwidth]{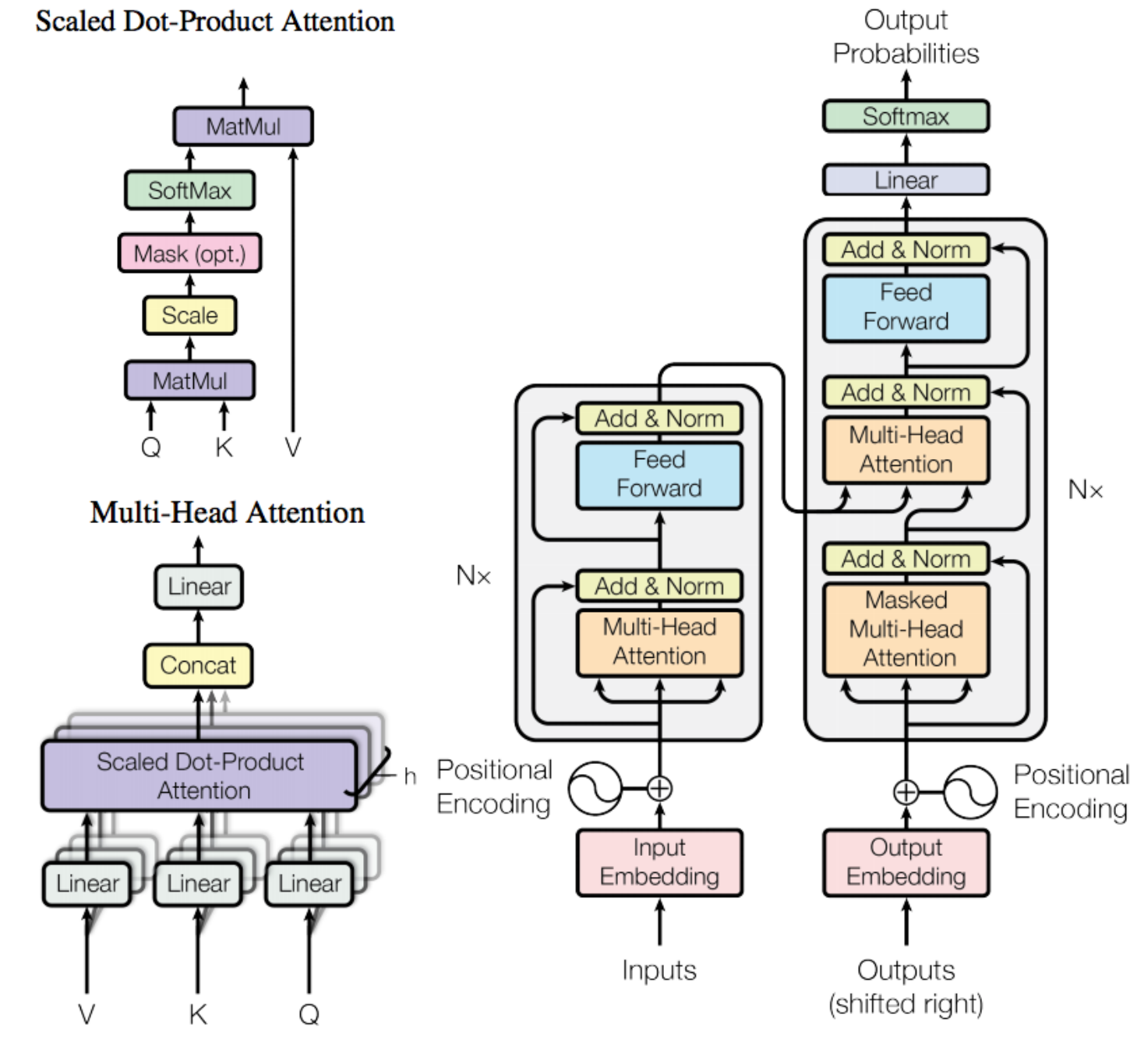}
\caption{The transformer model~\cite{vaswani2017attention} - an encoder-decoder model architecture and its key components.}
\label{fig:transformers}
\end{figure}

Self-attention is the key idea behind the transformer model.
It has the ability to attend to different positions of the input sequence to compute a representation of that sequence.
A transformer model handles variable-length input using stacks of self-attention layers instead of RNNs or CNNs.
An encoder-decoder model architecture is illustrated in Figure~\ref{fig:transformers}.

Multi-head attention consists of four parts, a set of linear layers that are split into heads, a scaled dot-product attention, a concat, and a final linear layer.
Each multi-head attention block gets three tensors as inputs; Q (query), K (key), V (value). 
These are put through the linear layers and split up into multiple heads.
The scaled dot-product attention computes the dot products of the query with all keys, and applies a Softmax function to obtain the weights on the values.
The attention output for each head is then concatenated and put through a final linear layer.
In addition to multi-head attention, each of the layers in our encoder and decoder contains a fully connected feed-forward network to improve the capacity of the model.
It consists of two linear transformations with activations in between.

By introducing parallelism on the sequence length,
dimensions of Q, K, and V tensors change unpredictably during serving.
Take a real-time translation system as an example.
After a short greeting phrase including a few words as input, a long paragraph of hundreds of words maybe its next input.
The variable-length input feature brings difficulties to batching. 
When the input contains a batch of request, in order to be processed together, the short sequence must be filled with zeros according to the longest sequence before being sent to the transformer models.

Figure~\ref{fig:transformers} shows a transformer architecture with both encoder and decoder parts.
Note that decoder parts are not necessary in transformer-based model, for example the widely-used BERT model only contain the encoder parts.

\subsection{Serving Systems}
Leveraging the power of transformer-based models to NLP online applications requires joint efforts in both training and serving stages.
Training is the process of building a model from offline data, which requires iteratively forward and backward passes.
Serving, consisting of repeated inferences, is the process of using the model to extract useful features from user online input through a forward pass.
Although inference does not involve complex backward computation and iterative processing,  its performance requirements are more demanding.
The serving process must run in real-time with low latency and sometimes should have the capacity of handling orders of magnitude more throughput than the training process.
A serving system can be divided into two components, a DL inference runtime, and a serving framework.

Training frameworks have been widely used as inference runtimes.
For instance, TF-serving~\cite{tfserving} is wrapped with TensorFlow.
Considering the poor performance of applying training framework in inference, 
some works have been dedicated to inference-specific runtimes, such as onnxruntime~\cite{onnxruntime}, 
TenorFlow XLA~\cite{xla}, TVM~\cite{chen2018tvm}, and TensorRT~\cite{tensorrt}.
Due to time-consuming preprocessing specific to the dimension of inputs, most of these runtimes can not be applied in variable-length input tasks.
Among them, only the onnxruntime is able to be used in the variable-length input tasks, with dynamic axis supports after version 1.3.
Faster Transformers~\cite{nvidiafaster} is a transformer boost software developed by NVIDIA.
However, it is not a complete runtime because it has no memory manager and has to be used as an operator for TensorFlow.
The comparison between this work and them are listed in Table ~\ref{fig:related-works}.

\begin{table}[ht!]
\scriptsize
\begin{centering}
\caption{Comparison of our runtime and existing GPU DL Inference runtimes.}
\label{fig:related-works}
\begin{tabular}{lcccccccc}
\hline
Related Works & Speed & Preprocess  &  Variable-Len  & Usage \\
\hline
Tensorflow-XLA~\cite{xla} & Medium & Yes & No & Easy \\
PyTorch~\cite{paszke2019pytorch} & Medium & No & Yes & Easy \\
TensorRT~\cite{tensorrt} & Fastest & Yes  & No  & Hard  \\
Faster Transformers~\cite{nvidiafaster} & Fast  & Yes  & No  & Hard  \\
ONNX-runtime~\cite{onnxruntime} & Fast & Yes  & Yes  & Medium  \\
\textbf{TurboTransformers} & \textbf{Fastest} & \textbf{No} & \textbf{Yes} & \textbf{Easy} \\    
\hline
\end{tabular}
\end{centering}
\end{table}

The serving framework wraps the runtime into a service exposing gRPC/HTTP as endpoints.
The advanced functionalities of the serving framework include batching, caching, model version management, and model ensembles.
Batching boosts throughput substantially by combining multiple inference requests into a batch to increase GPU usability, which is the main focus of this paper.
TF-serving enforces a static batch size by concat multiple requests together and has to pad zeros if requests are not enough.
Clipper~\cite{crankshaw2017clipper}  proposed an adaptive batching scheme to dynamically find and adapt the maximum batch size.
Nexus~\cite{shen2019nexus} further designed a batch scheduler to serve multiple different models on the same GPU.
Ebird~\cite{cui2019ebird} is a prototype system designed an elastic batch scheduler based on an inference engine supporting multiple batches of the same model running concurrently.
All of the above works are targeted at fixed-length input, which does not consider performance harm brought by zero-padding of short requests in a batch of variable-length requests.
To avoid the zero-padding problem in RNN, BatchMaker~\cite{gao2018low} breaks the computation graph into a graph of cellulars and dynamically decides the set of cellulars should be batched together.
It takes advantage of the weight sharing between multiple forward passes of RNN, which is not applicable in transformer models.


\section{Design Overview}
\label{sec:approach}
\begin{figure}[ht!]
\centering
\includegraphics[width=0.40\textwidth]{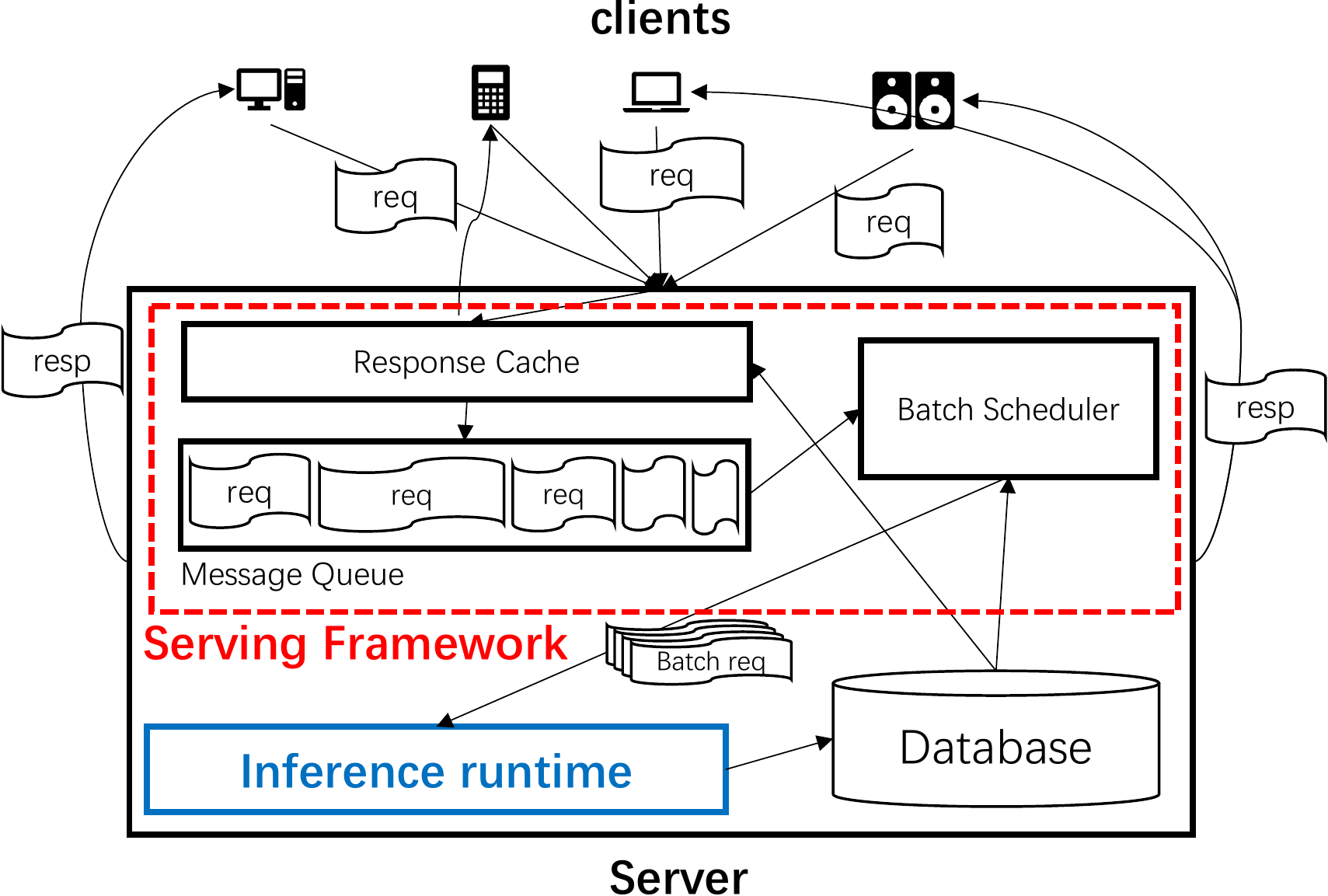}
\caption{The serving system architecture adopted by TurboTransformers.}
\label{fig:servering}
\end{figure}

As shown in the Figure \ref{fig:servering}, there are two ingredients of TurboTransformers, an \textbf{inference runtime} and a \textbf{serving framework}.
The system accepts network requests and responds the results processed by the Transformer models to end users.
Section~\ref{sec:runtime} will elaborate on the details of the runtime\footnote{The code of runtime is publicly available at \tiny{\url{https://github.com/Tencent/TurboTransformers}.}}, and Section~\ref{sec:framework} will focus on the serving framework.

\section{Inference Runtime}
\label{sec:runtime}
Inference runtime focuses on increasing computation efficiency and optimizing memory allocation. 

\subsection{Computational Optimizations}
\subsubsection{Kernel Fusion}
Without backward propagations, there is a lot of room left for inference customized optimizations.
When using training frameworks, like TensorFlow or PyTorch to infer a transformer model on the GPU, a significant amount of time is spent on some non-computational intensive kernels.
Take PyTorch as an example.
For the case where batch size and sequence length are both relatively large, PyTorch BERT brings low efficiency due to the inefficient implementations of non-GEMM kernels. 
For a BERT inference on a Tesla V100 GPU using input with batch size as 20 and sequence length as 128, 
only 61.8\% of the time is spent on GEMM kernels, and 38.2\% is spent on non-GEMM intensive cores, such as LayerNorm, Softmax, Add Bias, Transpose, etc.
For the case where batch size and sequence length are relatively small, PyTorch leads to poor efficiency due to the launch overhead of the CUDA kernels.
For a BERT model inference on a Tesla V100 GPU with batch size as 1 and sequence length as 40,
GPU is completely idle 80.64\% of the time.


\begin{figure}[ht!]
\centering
\includegraphics[width=0.5\textwidth]{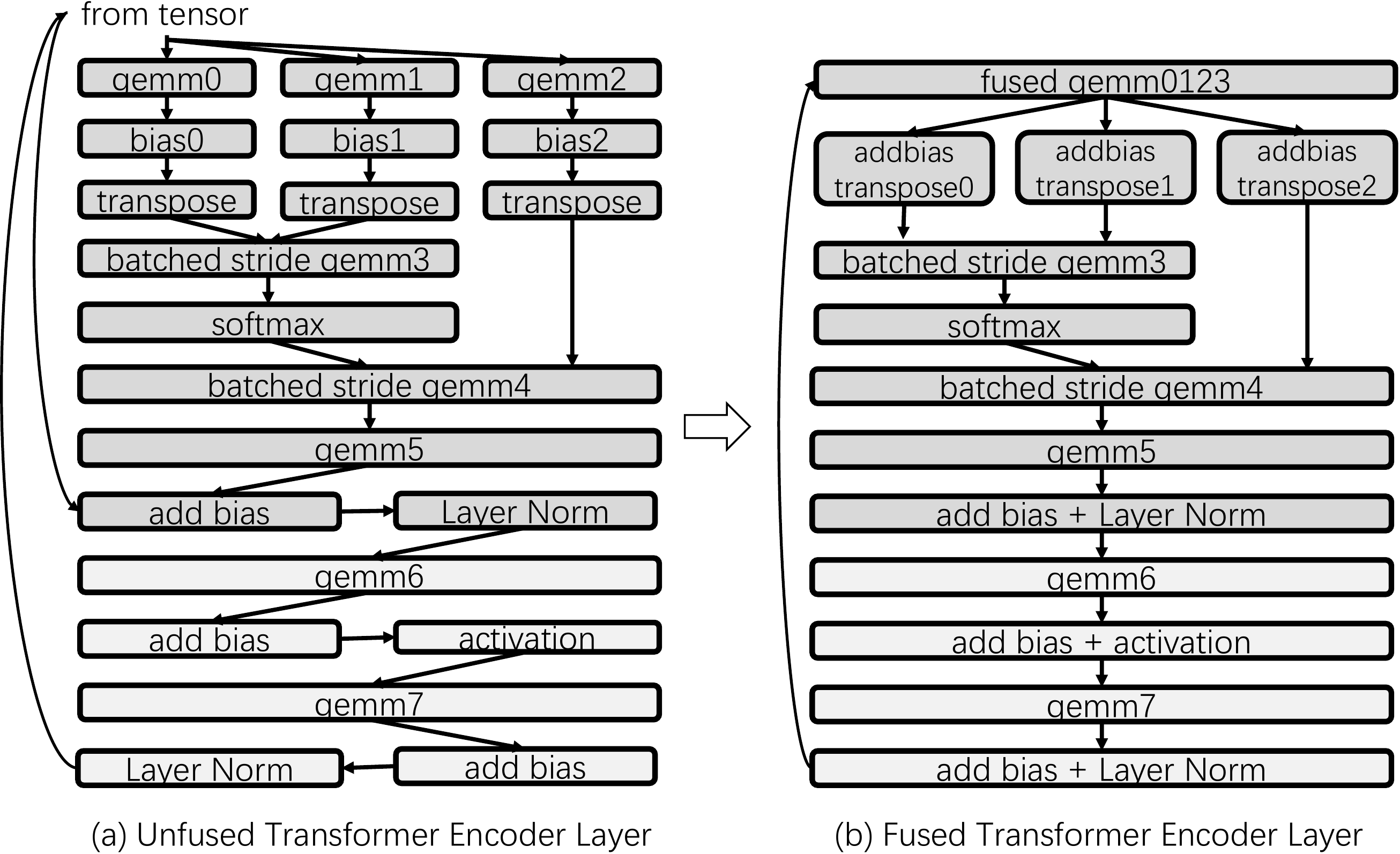}
\caption{Kernel fusion of a transformer encoder. The part in darker color is a multi-head attention. The part in lighter color is a feed forward network.}
\label{fig:kernel-fusion}
\end{figure}

Kernel fusion is able to increase computation efficiency by reducing the number of memory accesses, increasing cache locality, and reducing kernel launch overhead. 
Similar to many popular frameworks, such as TensorFlow, and Theano ~\cite{bergstra2010theano}, our runtime represents the DNN forward propagation by constructing a \textit{computation graph}, 
in which nodes are operators and edges are tensors.
As is shown in Figure \ref{fig:kernel-fusion}, the computation graph of a transformer can be reorganized into a more compact graph by fusing all the kernels between two GEMM kernels into a single one.
The fused non-GEMM kernels are non-standard DNN operators, so it is impossible to take advantage of existing DNN operator libraries, such as cuDNN~\cite{cuDNN}.
For example, there is no such API to combine matrix addition and transpose operation in a single CUDA kernel.

TurboTransformers implements all the non-GEMM kernels using CUDA, which can be categorized into two types.
One type of kernel, such as fused activation functions and fused transpose operations, are composed of element-wise operations.
There is no dependency between the processing of two different tensor elements, so we can process them in embarrassingly parallel.
The second type of kernels, including Softmax and fused LayerNorm, are composed of reduction operations, which is notorious for being hard to parallelized. 
The latter is the focus of our performance improvement.

%

\subsubsection{GPU-based Batch-Reduction}
\label{sec:gpu_batch_reduction}
Both of Softmax and LayerNorm based kernels can be viewed as \textit{Batch Reduction} operations.
On the lower dimension of a 2D tensor, Softmax calculates summation and maximum and LayerNorm calculates the mean and variance.
In other words, they both need to reduce a batch of 1D arrays in parallel.
Table~\ref{fig:batch-reduction-proportion} shows the proportion of time of the two operators in the attention layer.
In the table, the execution time of Softmax and LayerNorm is measured using PyTorch.
Attention time is measured using our runtime after replaced Softmax and LayerNorm with PyTorch's implementations, respectively.
We can observe that they are two big hotspots if not carefully optimized.

\begin{table}[ht!]
\tiny
\begin{centering}
\caption{Proportion of batch reduction operations in attention layer before and after optimizing.  }
\label{fig:batch-reduction-proportion}
\begin{tabular}{cc|cccccc}
\hline
\multicolumn{2}{c}{(batch size, seq len)}  & (1, 10) & (1, 100) & (1, 500) & (20, 10) & (20, 100) & (20, 500) \\
\hline
Softmax/ & before & 26.23\% & 24.73\% & 34.41\% & 3.04\% & 29.4\% & 90.68\% \\
Attention & after & 3.44\% & 3.18\% & 11.56\% & 2.46\% & 5.50\% & 15.46\% \\
\hline
LayerNorm/ & before  &  29.20\% & 21.72\% & 18.96\% &  10.61\% & 52.59\% & 83.38\% \\
Attention & after & 4.96\% & 4.40\% & 4.08\% & 5.14\% & 6.44\% & 4.24\% \\
\hline
\end{tabular}
\end{centering}
\end{table}

\begin{figure}[ht!]
\centering
\includegraphics[width=0.5\textwidth]{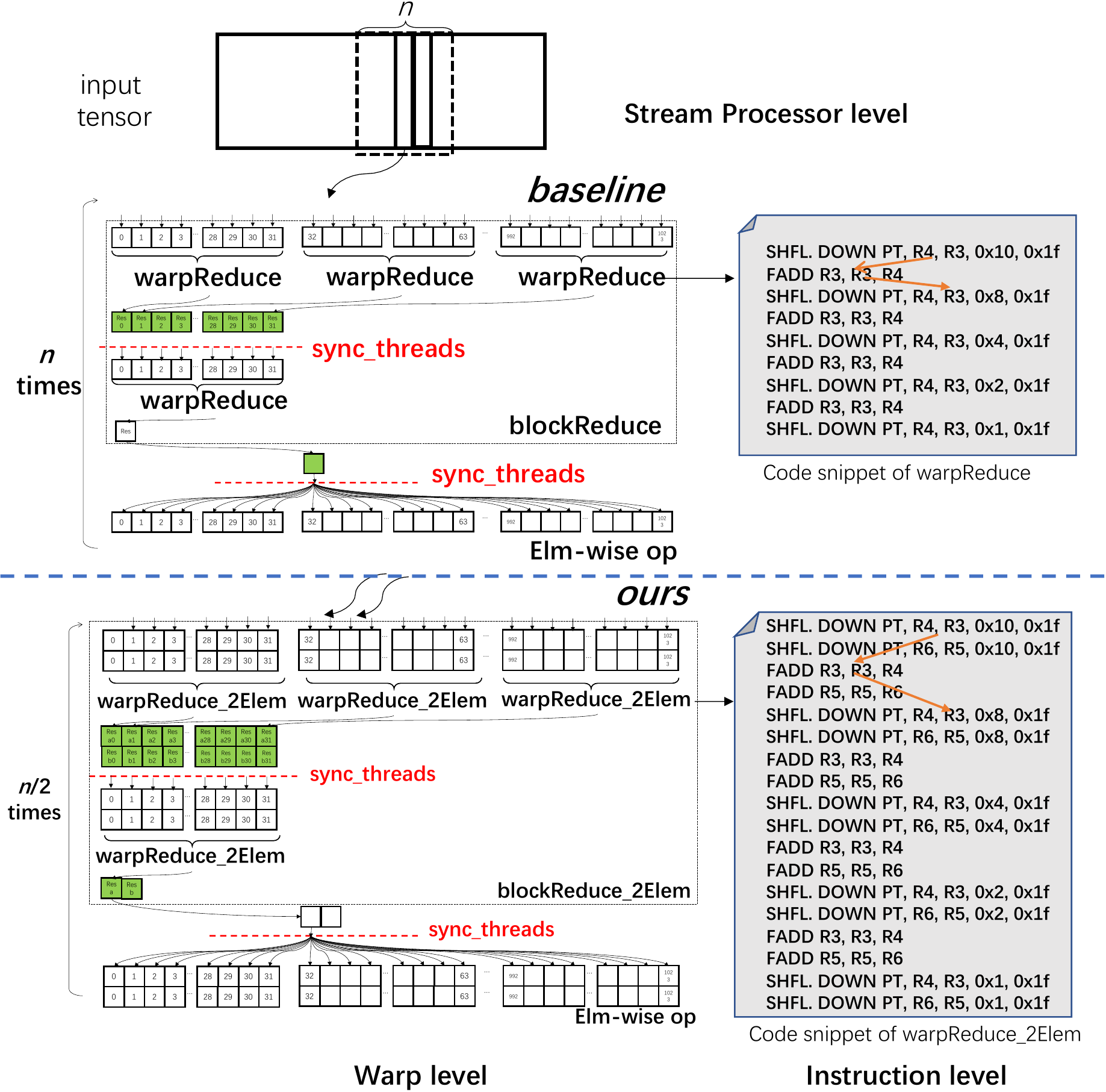}
\caption{Parallel batch reduction optimizations on three hardware levels.}
\label{fig:batch-reduction}
\end{figure}

Realizing the inefficiency of PyTorch operators, some effort is spent on optimize GPU batch reduction.
According to the programming model of CUDA, efficient reduction algorithms need to fully utilize the power of three hardware levels.
First, by splitting on the batch dimension, streaming-processors (SMs) level parallelism is exploited by the process of the workload on different SMs in parallel.
Second, warp level parallelism is exploited by taking advantage of the warp-level inter-thread communication mechanism provided by CUDA beyond the 9.0 version.
Third, instruction-level parallelism is exploited by overlapping memory access and computation instructions. 
A classical implementation adopted in Faster Transformers~\cite{nvidiafaster} is shown in the top part of the Figure~\ref{fig:batch-reduction}.
It is derived from work \cite{lin2019using},  which proposed a best practice for 1-D array reduction operations on the GPU.
Reduction workloads of $n$ rows (inside the dotted line on the top of the figure) are assigned to a thread block, which is scheduled to be executed on an SM.
The thread block sequentially performs $n$ times independent 1-D array reduction, which is finished with two-pass.
In the first pass, each warp of the SM conducts a warpReduce operation on 32 aligned elements, and then store reduction results of them inside shared memory.
In the second pass, a warp load at most 32 partial results to register and conduct another warpReduce to obtain the final results.

In this paper, we push the efficiency boundary of classical batch reduction algorithms on GPU.
Note that some space is still left for improvement of warp level and instruction parallelism in the above algorithm.
First, due to the accesses of shared memory, synchronizations of warps inside an SM introduce huge overheads.
Second, if the input array is not 32-aligned,
warp divergence resulting from boundary processing also introduces extra overhead.
Third, warpReduce leads to poor efficiency of instruction issuing.
As pointed out by reference \cite{lin2019using}, there is a dependency between shuffle and add instructions.
In the upper right corner of the Figure ~\ref{fig:batch-reduction}, the target register R3 in an SHFL DOWN instruction is required immediately as a source register in FADD instruction.
The FADD instruction can only be issued until the SHFL is completely finished, whose latency is more than 1 cycle.

The above three shortcomings can be overcome by leveraging parallelism between multiple 1-D reduction operations.
As shown in the bottom part of Figure~\ref{fig:batch-reduction},
a new subroutine warpAllReduceSum\_XElem (X = 2 in our figure) is introduced, which combine $X$ warp as a batch and do their reduction together.
First, only one synchronization is required for $X$ elements reduction in blockReduceSum\_XElem,
therefore, reduces $(X-1)/X$ synchronization cost.
Second, $X$ independent boundary processing can be merged into a single one, therefore reduce warp divergence.
Third, the instruction issuing is more efficient because we eliminate instructions dependency.
As shown in the figure, the target register of SHFL DOWN is required two cycles later by FADD as a source register.
Another SHFL DOWN  with no dependency on the previous one can be issued immediately.


\begin{figure}[ht!]
\begin{subfigure}
\centering
\includegraphics[width=0.5\textwidth]{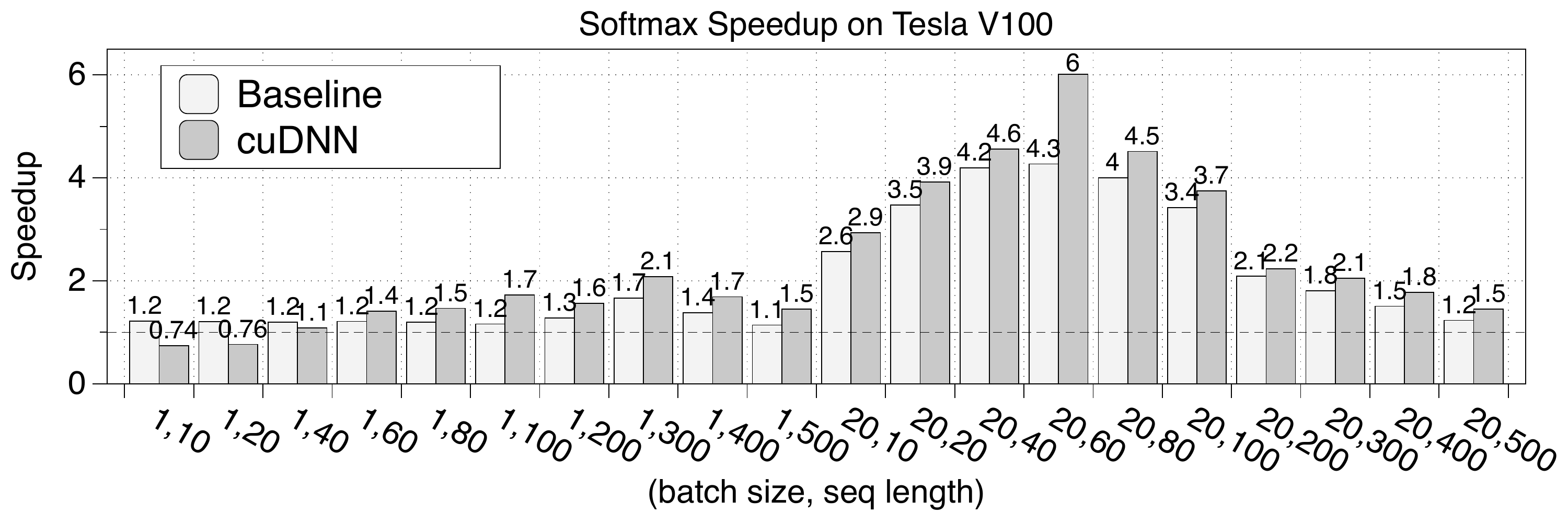}
\end{subfigure}
\begin{subfigure}
\centering
\includegraphics[width=0.5\textwidth]{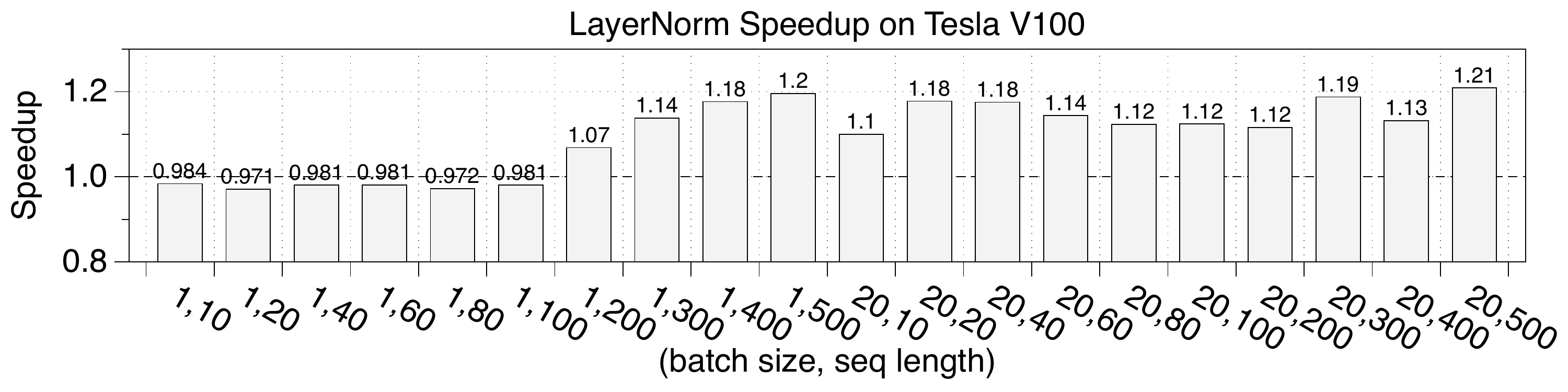}
\end{subfigure}
\caption{Speedup of batch reduction kernels on Tesla V100.}
\label{fig:batchreduction-speedup}
\end{figure}

Especially for the LayerNorm kernel, 
TurboTransformers further derives a mathematical optimization trick to improve efficiency.
LayerNorm requires the variances of elements of 1-D arrays.
There are two equivalent formulas of variance, as shown in the Equation \ref{seq:varience}. 
The first one used in ~\cite{nvidiafaster} requires two separate reductions for $x$ and $x-E (x)$, during which one synchronization between.
We use the second one to compute variances. 
The warpAllReduceSum\_2Elem can simultaneously reduce $x$ and $x^2$, which not only increases the efficiency of instruction execution but also reduces half of the number of synchronizations.

\begin{equation}
\label{seq:varience}
\text{Var}(x) = E(x - E(x))^2 = E(x^2) - E^2(x)
\end{equation}

Figure ~\ref{fig:batchreduction-speedup} shows the speedups of Softmax and LayerNorm kernels in TurboTransformers compared with the other implementations. 
The baseline of both kernels is used implementations from ~\cite{nvidiafaster}.
For the Softmax kernel, we also compare with the Softmax routine cuDNNv7.5.
In most cases, our optimization strategy has achieved obvious acceleration.
Due to the batch dimension is larger for Softmax, its performance boost is more significant.

%

\subsection{Memory Manager}
\label{sec:memory_manager}
Besides computing optimizations, memory management is also vital to a DNN runtime.
It is evaluated by two metrics.
The \textit{allocation efficiency}, which determined by the number of times and the amount of the memory is allocated and released, affects the overall execution speed of the runtime.
The \textit{memory footprint} affects the possible size of the model as well as the maximum batch size of requests.
Three types of memory are managed by the runtime, i.e., input tensors, intermediate tensors, layer parameters.
Dimensions of intermediate tensors change frequently during inferences in case of variable-length input.
Adversely, for fixed-length input, the dimensions of intermediate tensors are determined in advance and allocation scheme never changes during inference processes.
Therefore, the memory usage and positions of tensors can be pre-optimized and fixed during serving.
However, The optimal memory allocation strategy is different in case of the different input lengths.
Allocator of variable-length input must efficiently deal with unpredictable memory usage requirements.

It is complicated to achieve both a high allocation efficiency and a small memory footprint for variable-length input.
To achieve the best memory footprint, the memory of intermediate tensors should be allocated when needed, and released immediately when not required.
Frequent allocation and release of small memory space on GPU will lead to a worse runtime efficiency.
For example, in this case, 50\% of the computing resources idle wait for memory allocation ready on Tesla M40 (batch size = 20, sequence length = 128).
To achieve the best allocation efficiency,
the memory should be allocated in advance and cached for repeated use in the future the inferences without any extra allocation.
However, it is challenging to predict maximum memory usage to meet the requirement of a long-term serving process.
Even if we know it, occupying a vast amount of memory for a long time will lead to extremely poor memory footprints.

Although the allocator for fixed-length input has been well studied, the one for variable-length input is still not perfect.
For fixed-length input, by taking advantage of the topology of the computation graph, works ~\cite{pisarchyk2020efficient}~\cite{lee2019device} figure out the optimal memory usage by reusing the same memory space for intermediate tensors that do not coexist.
For variable-length input, The existing solutions have to tradeoff allocation efficiency and footprint.
PyTorch~\cite{paszke2019pytorch} designed a custom caching tensor allocator
which incrementally builds up a cache of CUDA memory and reassigns it to later allocations.
PaddlePaddle~\cite{ma2019paddlepaddle} used a similar method, which is all inspired by the caching device allocator implemented in the NVlab's cub library~\cite{nvlab}.
Experiments have shown that these methods cannot achieve optimal memory usage,
because they do not consider the DNN's computation graph.

\begin{algorithm} \SetAlgoLined
\caption{Sequence-length-aware allocator}
\label{algo:allocator}
\small
    \SetKwProg{Def}{def}{:}{end}
    \Def{FindGapFromchunk($t : tensor\_id$,  $chunk$)}{
        get $chunk\_size$ from $chunk$;  $smallest\_gap \leftarrow \infty$\;
        $prev\_offset \leftarrow 0$; $best\_offset \leftarrow NIL$\;
        \ForEach { record $x \in $ ${chunk}$} {
        		$max\_first\_op \leftarrow$ max ($first\_op_t$, $first\_op_x$) \label{algo:findgap:ref1}\;
    		$min\_last\_op \leftarrow$ min ($last\_op_t$, $last\_op_x$)\;
    		\If {${max\_first\_op \leq min\_last\_op}$} {  \label{algo:findgap:ref2} 
    			$gap \leftarrow offset_x - prev\_offset$ \label{algo:findgap:ref3}\;
    			\If {$gap \geq size_t$ and $gap < smallest\_gap$} {
    				$smallest\_gap \leftarrow gap$; $best\_offset \leftarrow prev\_offset$\; 
    			} \label{algo:findgap:ref4}
    			$prev\_offset \leftarrow$ max($prev\_offset, offset_x + size_x$)\;
    		}
        }
        \If {$best\_offset$ is $NIL$ and $chunk\_size - prev\_offset \geq size_t$} { \label{algo:findgap:ref5} 
        		$best\_offset \leftarrow prev\_offset$\;
       }
       \If {$best\_offset$ is $NIL$} {
    		\Return {${INVALID}$}
       }
       \Return {${best\_offset}$}  \label{algo:findgap:ref6}
    }
    \SetKwProg{Def}{def}{:}{}
   \Def{MemAllocate({$tensor\_usage\_records$ : \textit{a list of tuples (first\_op, last\_op, size)}, $chunk\_list$ : \textit{a chunk has size, mem addr, list of <tensor\_id, offset>}})}{
        {sort $tensor\_usage\_records$ in decreasing order of $size$}\;
        \ForEach {record $t$ $\in$ $tensor\_usage\_ records$} {
        		${is\_assigned \leftarrow false}$\;
    		\ForEach {${chunk \in chunk\_list}$} {
    			${offset}$ $\leftarrow$ $\textbf{FindGapFromchunk}$($t$, $chunk$)\;
    			\If {${offset}$ is valid} {
    				${assigned\_chunk_t} \leftarrow chunk\_id$; ${assigned\_offset_t} \leftarrow offset$\;
    				${is\_assigned \leftarrow True}$\; 
				$\textbf{break}$\;
    			}
    		}
    		\If {${is\_assigned}$ is false} {
    			${new\_chunk\_size \leftarrow }$ max(DEFAULT\_CHUNK\_SIZE, ${t\_size} \times$ K\_SCALE)\;
    			${assigned\_chunk_t} \leftarrow$ len$(chunk\_list)$; ${assigned\_offset_t} \leftarrow 0$\;
    			append a new chunk of size $new\_chunk\_size$ to $chunk\_list$\;
    		}
        }
        release unused chunk in $chunk\_list$\;
        \Return {${assigned\_chunk}$, ${assigned\_offset}$, $chunk\_list$}\;
    }
\end{algorithm}

We adopted an innovative memory optimization method for variable-dimension intermediate tensors, which evoke a light-weight variable-length-aware memory manager after knowing the length of each inference.
To achieve more efficiency and less footprint, the allocator used in TurboTransformers combines the idea of memory cache and graph-topology-aware space reuse.
First, it organizes memory space in units of the chunks, which is a small block, for example, 2MB of memory.
By reusing already allocated chunks, allocation efficiency can remain at a high level during serving.
Second, it utilizes the computation graph to know the life cycle of each intermediate tensor in advance,
and calculate the offset of each tensor within a specific chunk as soon as it recognizes the sequence length of the new arrival request.
In this way, tensors with no overlapping life cycle can reuse the same memory space, therefore reduce memory footprint as much as possible.

Our sequence-length-aware allocation method is shown as Algorithm \ref{algo:allocator}.
The input tensor usage record is a list of tuples \{$first\_op$, $last\_op$, $size$\}, where $first\_op$, $last\_op$ are indices of the first and last operator that use the tensor.
The indices are from the topological sorting of the DNN's computation graph.
It first sorts the usage records in non-increasing order based on the size of the tensor.
$FindGapFromchunk$ is used to determine if there exists a free gap inside a chunk to fit for that tensor.
If no such gap is found in all existing chunks, we append a new chunk to the end of the chunk list.
The size of the new chunk is the maximal one of DEFAULT\_chunk\_SIZE (2MB in our implementation) and the size of tensor times K\_SCALE (1.2 in our implementation).
When a chunk is not used in this inference, in our algorithm, its memory is released immediately.
Alternatively, we can assign each chunk a maximum inference idle times, and release it after it reaches the time limit.


$FindGapFromchunk$  of Algorithm \ref{algo:allocator} finds the best gap in a memory chunk.
It is equivalent to a special case of the 2D strip packing problem, which is NP-hard.
We slightly modify the Greedy by Size for Offset Calculation Algorithm in ~\cite{pisarchyk2020efficient} to solve FindGapFromchunk in $O(n^2)$ time complexity.
The inputs are a target tensor $t$ and a target chunk $chunk$.
For each record $x$ in the chunk, the algorithm first check whether $x$ and the target tensor $t$ overlap in usage time (L\ref{algo:findgap:ref1}-L\ref{algo:findgap:ref2}),
in order to find the smallest gap between them such that current tensor fits into that gap (L\ref{algo:findgap:ref3} - L\ref{algo:findgap:ref4}).
If we found such a gap before the end of the chunk, we assign the tensor $t$ to the gap.
Otherwise, the function return invalid. (L\ref{algo:findgap:ref5} - L\ref{algo:findgap:ref6}).

\begin{figure}[ht!]
\centering
\includegraphics[width=0.5\textwidth]{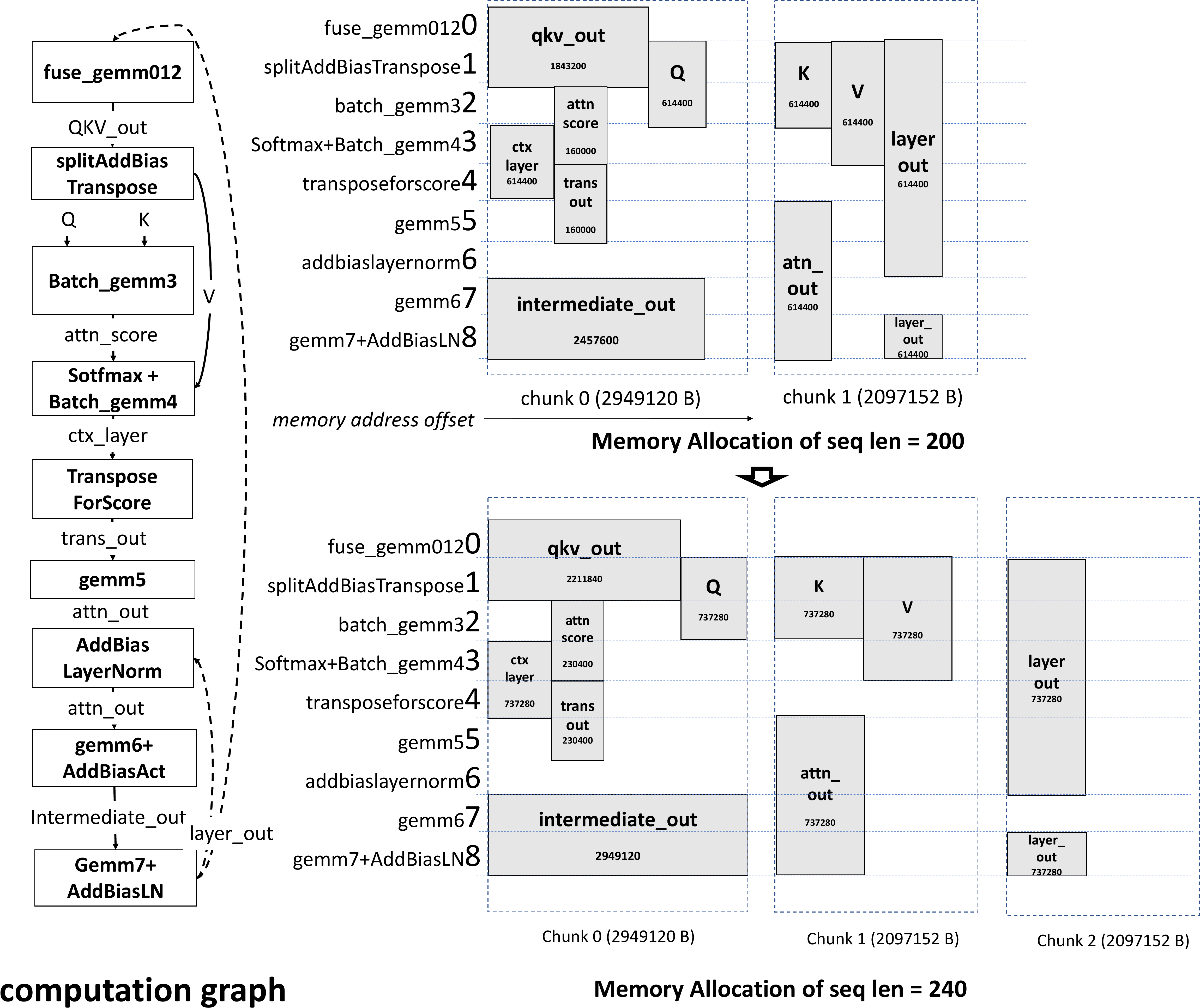} 
\caption{A memory allocation example uses our proposed variable-length-aware allocator.}
\label{fig:memory-allocation-example}
\end{figure}

Figure \ref{fig:memory-allocation-example} shows an example of applying our algorithm on a BERT inference application.
When the input length changes from 200 to 240, we allocate one more chunk and adjust the offsets.

 
\section{Serving Framework}
\label{sec:framework}
\begin{figure}[ht!]
\centering
\includegraphics[width=0.5\textwidth]{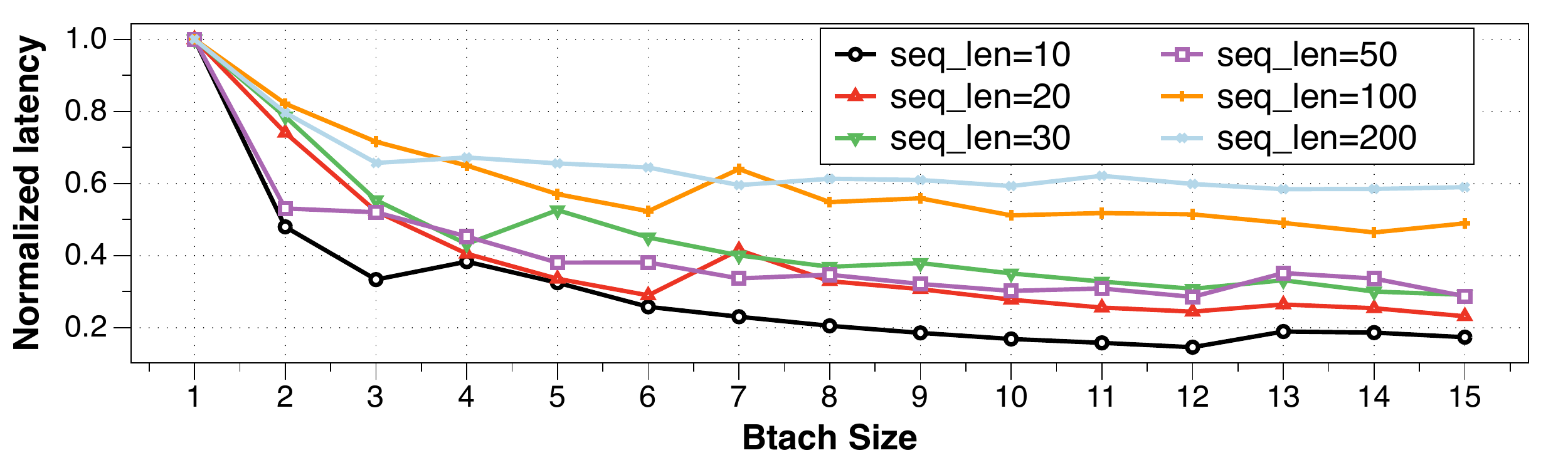}
\caption{Batching brings performance gain for the base BERT serving on RTX 2060 GPU. The y-axis illustrates the normalized latency of inferring a request of batch size 1.}
\label{fig:serving-batch-speedup}
\end{figure}

Based on the computation runtime, a serving framework is required to attain enough serving throughput under latency constraints to satisfy a Service Level Objectives (SLOs).
The serving framework of TurboTransformers is shown in Figure ~\ref{fig:servering}.
The user's requests first arrive at a \textit{Message Queue} (MQ) and then are sent to runtime for inference computation after two serving-level optimizations, i.e. \textit{Caching} and \textit{Batching}. 
For caching, similar to Clipper~\cite{crankshaw2017clipper}, by caching the inference results in a database, the \textit{Resp Cache} component in the figure responses the frequent requests without evaluating the model.
For batching, the \textit{Batch Scheduler} component is responsible for packaging requests that come in a period of time into a batch.
In most application scenarios, the input of the user is a single inference request with batch size as 1.
It has been known that small batch sizes lead to low GPU hardware utilization.
Packaging multiple requests into a relatively larger batch and conducting inference on them together can improve hardware utilization.
As shown in Figure~\ref{fig:serving-batch-speedup}, serving requests in batch brings significant speedup, especially for short sequences.
The batch schedulers adopted by conventional serving systems, like TF-serving~\cite{tfserving} and Nexus~\cite{shen2019nexus}, are designed to packages requests with fixed-length into a batch.
Currently, serving systems are lack of critical ability to handle variable-length requests.


\begin{figure}[ht!]
\centering
\includegraphics[width=0.5\textwidth]{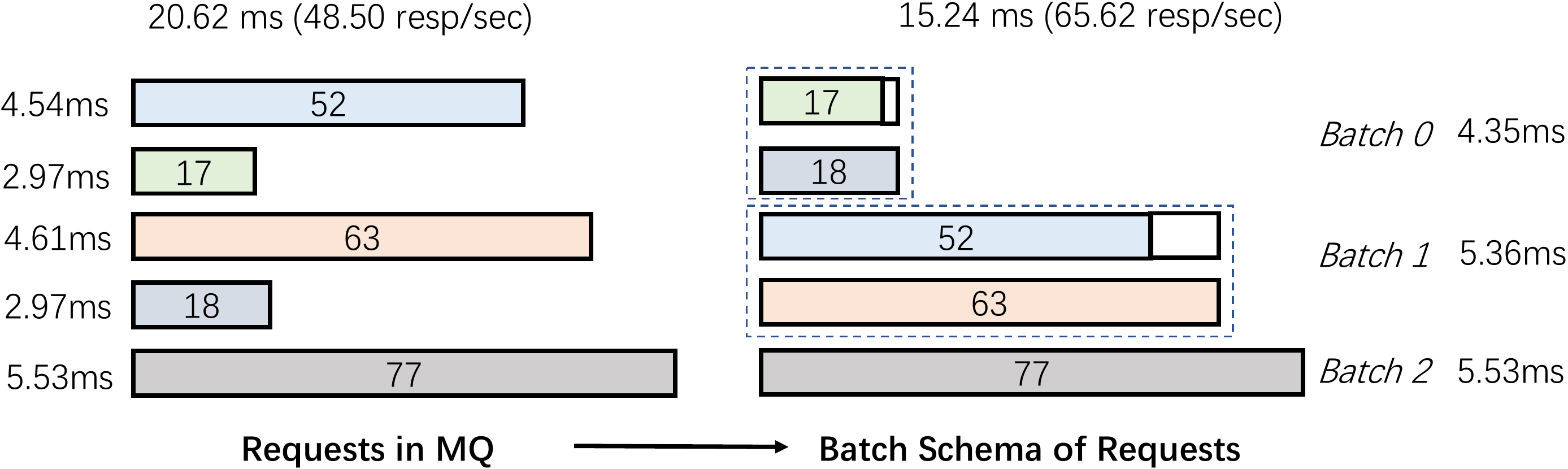}
\caption{An example of batch scheduler for variable-length requests.}
\label{fig:batch-example}
\end{figure}

How to batch variable-length requests to achieve the optimal throughput is tricky.
If we package multiple requests of different lengths into a batch, then all requests in the batch will be zero-padded with regards to the maximum length of the request in the batch.
Zero paddings will introduce a lot of extra computations. 
An efficient batch scheduler has to carefully balance the overheads of padding and the benefits of batching.
For example, assume that there are five inference requests to be served, with lengths of 17, 18, 52, 63, and 77, respectively.
Packing a single batch with 5 instances is less efficient than no batching.
The batching scheme achieving the optimal throughput is packing three batches.
As shown in Figure ~\ref{fig:batch-example}, the response throughput (resp/sec) improved 35\% by the optimal scheduling scheme.

For variable-length request serving, we proposed an innovative sequence-length-aware batch scheduler in Algorithm~\ref{algo:batchscheduler}.
The core idea is to use dynamic programming (DP) to solve an optimization problem which maximizes the response throughput as the objective in $O(n^2)$ time complexity.
The inputs of the algorithm consist of two data structures.
The ${request\_list}$ is the list of input requests with variable length.
The ${cached\_cost}$ is a dictionary.
It uses two keys as indices, namely the sequence length and batch size.
Its value is the inference cost with the parameter from the corresponding key.
The values of $cached\_cost$ are collected by a warm-up phase after the service first starts on specific hardware, 
which utilizes the runtime to run inferences under all possible batch sizes and sequence lengths.
They are stored on disk or database (database in Figure ~\ref{fig:servering}) and reloaded to memory when the serving module is restarted. 
First, the $request\_list$ are sorted in increasing order according to the sequence length.
Then line 3 initializes an array called $states$ to store intermediate information.
Specifically, $state[i]$ records the minimum time overhead for processing the sublist $request\_list[0:i]$.
The algorithm traverses each request in $request\_list$ and uses a DP algorithm to update the $state[i]$ at the corresponding position $i$.
The Bellman equation of this DP problem is shown in Equation \ref{seq:dp}.

\begin{equation}
\begin{aligned}
\label{seq:dp}
state[i] = \mathop{min}_{0\leq j \leq i}(cached\_cost[request\_list[i].len][i-j+1] \\
\times(i-j+1) + states[j-1])
\end{aligned}
\end{equation}

\begin{algorithm}
\small
\caption{Batch Scheduler With DP}
\label{algo:batchscheduler}
    \KwIn{ ${request\_list}$, ${cached\_cost}$} 
    {sort $request\_list$ in increasing order with regards to sequence length}\;
    $N \leftarrow $ Size($request\_list$)\;
    Create $states, start\_idx\_lis$ as lists of size $N+1$\;
    $states[0] \leftarrow 0$; $i \leftarrow 1$\;
    \While{$i \leq N$} {
		$j \leftarrow i - 1$; $start\_idx = i - 1$\;
		$cur\_length = request\_list[i - 1].length$\;
		$min\_cost = cached\_cost[cur\_length][1] + states[j]$\;
		\While{$j > 0$} {
			 $tmp\_cost = states[j-1] + cached\_cost[cur\_length][i - j + 1]*(i - j + 1)$\;
			\If{$tmp\_cost < min\_cost$} {
				{$min\_cost = tmp\_cost$}; $start\_idx = j - 1$\;
			}
			$j \leftarrow j -1$\;
		}
		$states[i] = min\_cost$; $start\_idx\_list[i] = start\_idx$\;
		$i \leftarrow i + 1$\;
	}
	$i = N$\;
	 \While {$i > 0$} {
	 	$end\_idx \leftarrow i$; $start\_idx \leftarrow start\_idx\_list[i]$\;
		pack ${request\_list[start\_idx : end\_idx]}$ into a batch\;
		$i = start\_idx- 1$\;
	 }
\end{algorithm}

Note that the premise of this algorithm work is that there exists a scheduling strategy for requests in MQ that can meet SLO on this server.
In a multi-server environment, an upper-level load balancer as the one in Nexus~\cite{shen2019nexus} can ensure that the requests assigned to each server will not be overloaded.
There are two options to decide when to evoke the batch scheduler.
The first one is a hungry strategy.
When the runtime is idle, we immediately start the batch scheduler to batch requests in MQ.
This strategy is suitable for the situation where request throughput is high and the GPU have to run at full load.
The second one is a lazy strategy.
Similar to delayed batching of Clipper, we sets a timeout value and a maximum batch size.
Once the number of requests in the batch exceeds the maximum batch size, or the timeout is reached, we start the batch scheduler.
Due to the reordering of requests in Algorithm ~\ref{algo:batchscheduler} , requests that arrive early may be served late.
We check the elapse between the current time and the recorded arrival timestamp of request at the front of the MQ,
and start the batch scheduler immediately if the elapse plus the estimated execution latency of current requests in batch exceeds half of the latency constraints.


\section{Experimental Results}
We evaluated the performance of both the runtime and the serving system on a server equipped with an AMD Ryzen 7 3700X CPU and one RTX 2060 GPU.

\subsection{Usability}
Our runtime provides C++ and Python APIs.
For the Python one, as shown in the following code snippet, adding 3 lines of python code (L3, 12, 13) can bring end-to-end speedup.

\begin{lstlisting}[language=Python, basicstyle = \small]
import torch
import transformers
import turbo_transformers
model_id = "bert-base-uncased"
torch_model = transformers.BertModel.from_pretrained(model_id)
torch_model.eval()
input_ids = torch.tensor(
        [12166, 10699, 16752, 4454],
        dtype=torch.long, device = torch.device('cuda:0'))
torch_model.to(torch.device('cuda:0'))
torch_res = torch_model(input_ids)
turbo_model = turbo_transformers.BertModel.from_torch(torch_model)
turbo_res = turbo_model(input_ids)
\end{lstlisting}

\subsection{Performance of the Runtime}
The runtime is evaluated on four transformer DNNs, including Bert~\cite{devlin2018bert}, Albert~\cite{lan2019albert}, DistilBert~\cite{sanh2019distilbert} and a Seq2Seq Decoder~\cite{vaswani2017attention}, the parameters of which are shown in Table ~\ref{tab:expriment-parameter}.
The former three DNNs consist of Transformer encoder structures, and the latter consists of both encoder and decoder structure and is applied in a Neural Machine Translation system.
The Bert model adopts a base configuration, while the Albert model adopts a large configuration.

\begin{table}[ht!]
\scriptsize
\begin{centering}
\caption{Evaluated transformer models \& parameters}
\label{tab:expriment-parameter}
\begin{tabular}{lclclclc|c|c|c|cl}
\hline
Model & Parameters \\
\hline
Bert & num\_layer=12, num\_head=12, hidden\_size=4096, inter\_size=3072 \\
\hline
Albert & num\_layer=12, num\_head=64, hidden\_size=4096, inter\_size=16384\\
\hline
DistilBert & num\_layer=6, num\_head=12, hidden\_size=4096, inter\_size=3072\\
\hline
Seq2Seq & num\_layer=6, num\_head=16, hidden\_size=3072,  \\
Decoder & beam\_size=4, max\_target\_len=500 \\
\hline
\end{tabular}
\end{centering}
\end{table}

\subsubsection{End-to-end Speed on Variable-length Input}

\begin{figure}[ht!]
\begin{subfigure}
\centering
\includegraphics[width=0.5\textwidth]{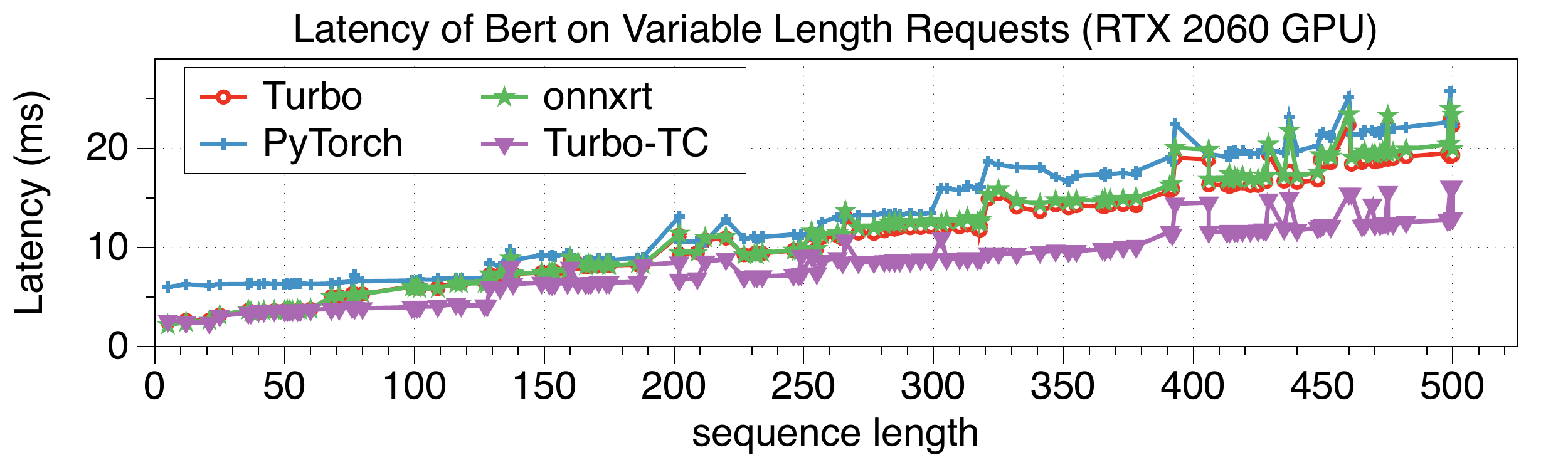}
\label{fig:bert-latency}
\end{subfigure}
\begin{subfigure}
\centering
\includegraphics[width=0.5\textwidth]{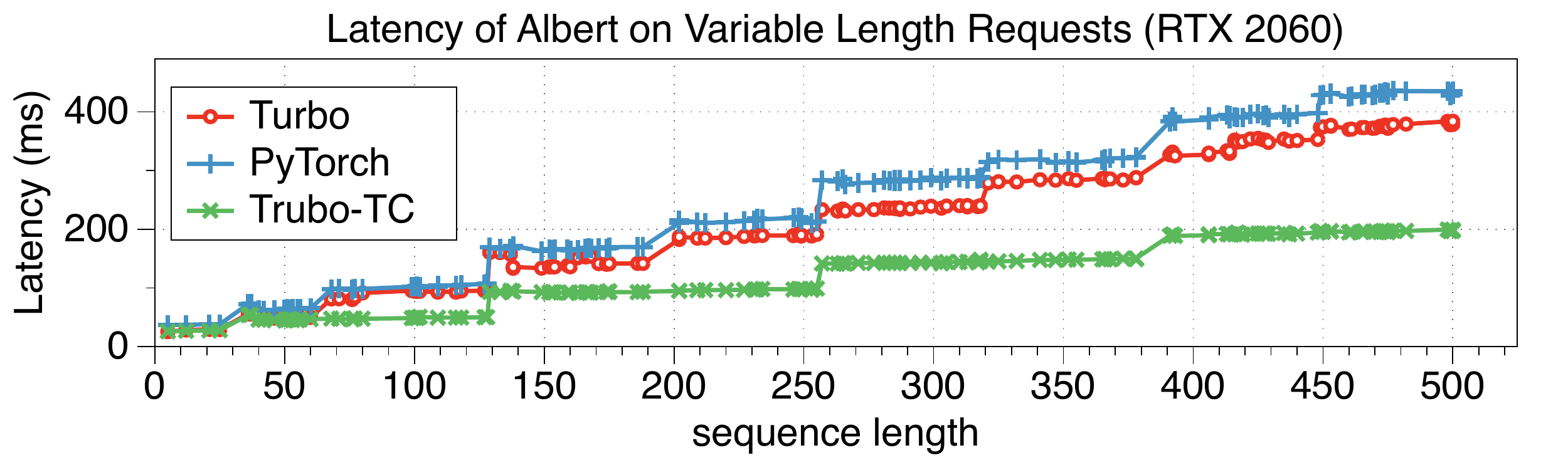}
\label{fig:albert-latency}
\end{subfigure}
\begin{subfigure}
\centering
\includegraphics[width=0.5\textwidth]{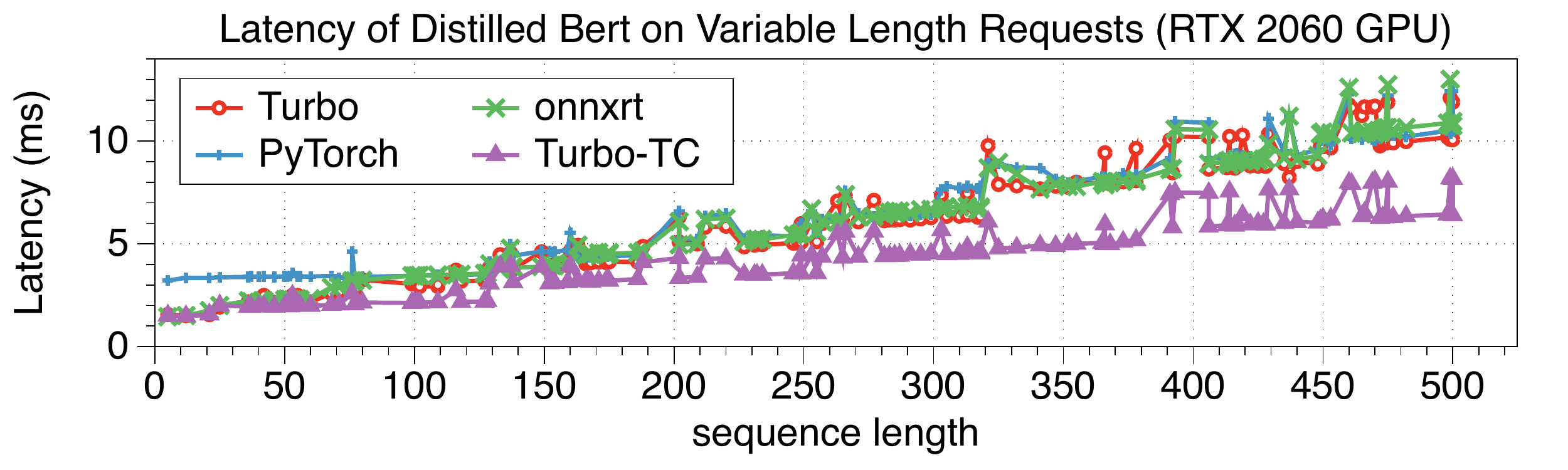}
\label{fig:distilbert-latency}
\end{subfigure}
\begin{subfigure}
\centering
\includegraphics[width=0.5\textwidth]{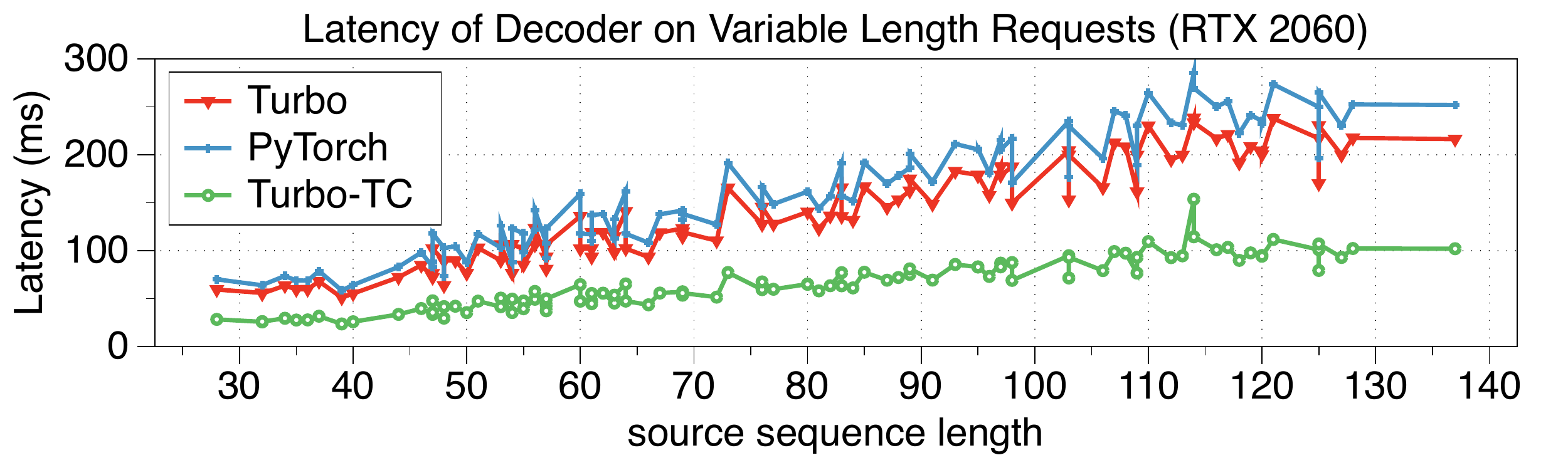}
\label{fig:decoder-latency}
\end{subfigure}
\caption{Benchmarking the latency of runtimes using variable length requests.}
\label{fig:variable-latency}
\end{figure}

The ability to handle variable-length input is evaluated by sequential execution of the runtime using requests of different lengths.
For Bert and Albert, the input requests are randomly generated texts whose lengths are uniformly distributed from 5 to 500.
For Decoder, it is a model adopted from a Chinese-English translation task, and its inputs are randomly sampled Chinese texts whose length ranges from 28 to 137.
The performance results are shown in Figure ~\ref{fig:variable-latency}.
Note that our generating and sampling processes are completely random, and the random seed is the same for different tests,
although the figure displays in order of input length from small to large for the sake of clearness.
We compare our runtime with the PyTorch (v1.5.0) and onnxruntime-gpu (v1.3.0).
Turbo is the performance of our proposed runtime implemented by FP32 GEMM algorithms.
The turbo-TC allows GEMM operation to use Tensor Core~\cite{markidis2018nvidia} if possible.
Because the tensor core optimization is not allowed in PyTorch and onnxruntime, we list it here as an additional reference. However, it introduces minimal and acceptable precision loss to the FP32 version.

TurboTransformers' runtime shows significant performance advantages over PyTorch on short requests.
In the case of Bert inference, Turbo's speedups to PyTorch are ranging from 0.97x-2.44x, on average 1.25x.
In Albert inference, Turbo's speedups to PyTorch are ranging from 1.04x-1.45x, on average 1.17x.
In the case of DistilBert inference, Turbo's speedups to PyTorch are ranging from 0.85x-2.24x, on average 1.13x.
In the case of Decoder inference, Turbo's speedups to PyTorch are ranging from 1.14x-1.20x, on average 1.16x.
The performance improvement is more obvious in test cases of short input sequences.
The latency of long input sequences is mainly limited by GEMM operations, which are not optimized by our runtime.
Since the GEMM operations are conducted on the tensor core, Turbo-TC always leads to much low latency.
For the same reason, the decrease in Albert's speedup to Bert is caused by the larger parameters of the Albert model, leading to an increase in the proportion of GEMM operations.

TurboTransformers exhibits similar performance with onnxruntime.
In the case of Bert, the speedups of Turbo to onnxruntime are ranging from 0.88x-1.05x, on average 1.01x.
In the case of DistilBert, the speedups of Turbo to onnxruntime are ranging from 0.83x-1.36x, on average  1.03x.

\begin{figure}[ht!]
\centering
\includegraphics[width=0.5\textwidth]{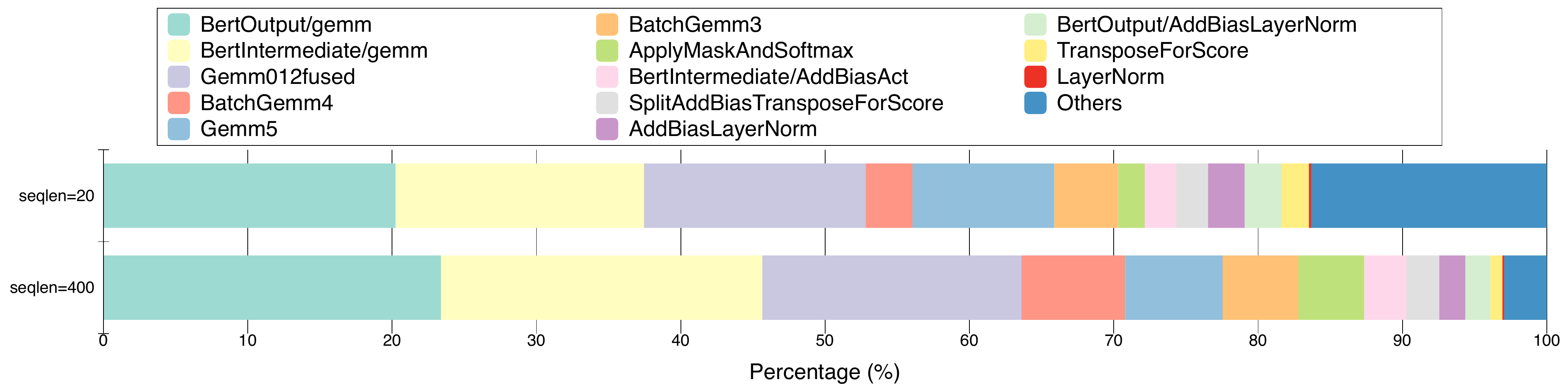}
\caption{Time distribution of different BERT computation kernels in the case of long and short sequences.} 
\label{fig:TimeDistGPU}
\end{figure}

The time distribution of different BERT computation kernels is analyzed in Figure~\ref{fig:TimeDistGPU}.
We selected two test cases to show the inference hotspots in a long sequence input as 400 and a short sequence input as 20.
In the sequence length 20, the GEMM kernels account for 70.31\% of overall computation time.
The Softmax kernel (ApplyMaskAndSoftmax) only accounts for 1.85\% of time, and LayerNorm kernels (AddBiasLayerNorm + BertOutput/AddBiasLayerNorm) account for 2.71\% of time.
In the sequence length 400, the GEMM kernels account for 82.80\% of overall computation time. 
The Softmax accounts for 4.57\%, and LayerNorm accounts for 3.64\%.
The remaining time is spent in element-wise operations, such as activations, biases addition, tensor transposing, and reshaping.
Our optimization for batch reduction is very successful since they are no longer the main hotspots among the non-GEMM kernels.

%
%
%

\subsubsection{Memory Optimization on Variable-length Requests}
\begin{figure}[ht!]
\centering
\includegraphics[width=0.5\textwidth]{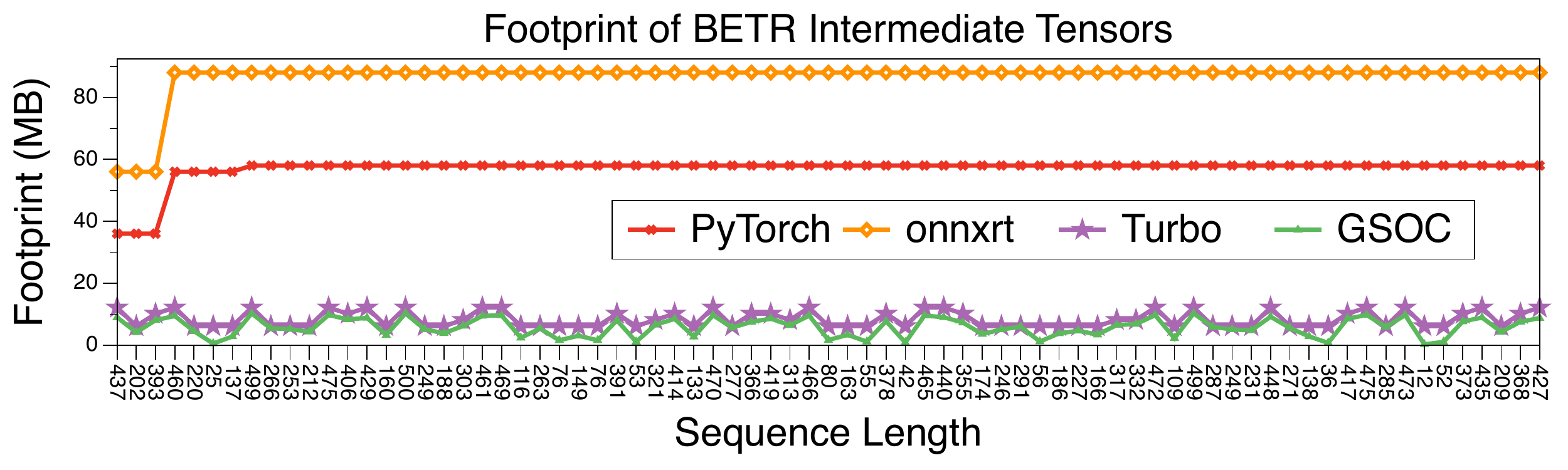}
\caption{Footprint of intermediate tensors during BERT inferences.}
\label{fig:intermediatefootprint}
\end{figure}

\begin{figure}[ht!]
\centering
\includegraphics[width=0.5\textwidth]{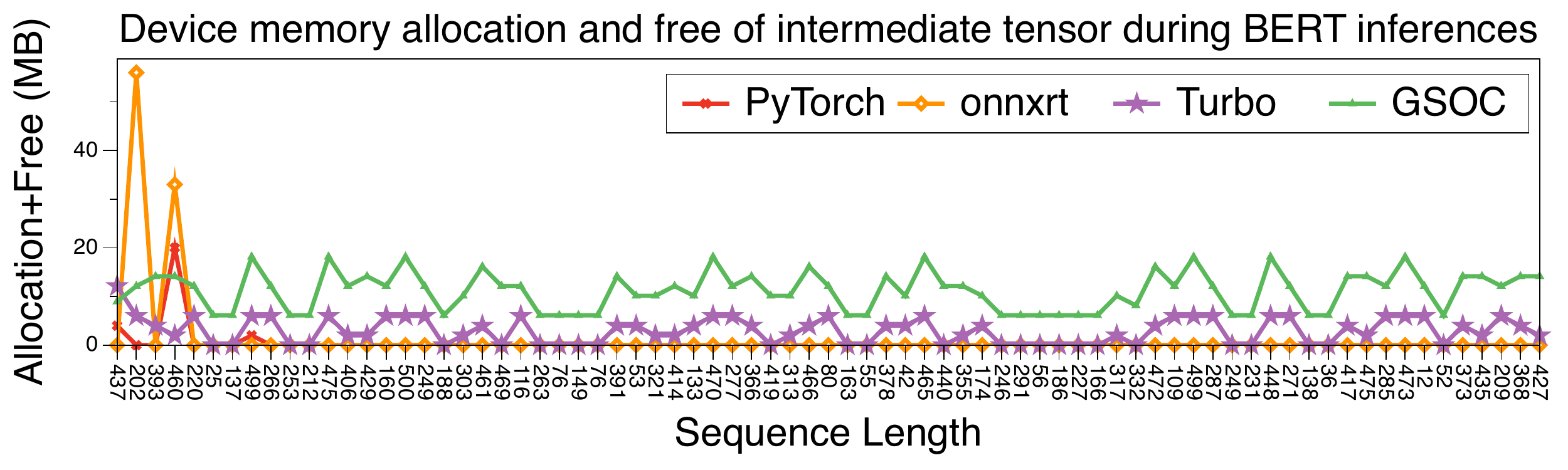}
\caption{Amount of device memory allocation/free for intermediate tensors during BERT inferences.}
\label{fig:allocateintermeddiat}
\end{figure}

For memory optimization, we first analyze the memory footprint of the intermediate tensors.
Our proposed model-aware-allocator is compared with three other allocators,
including the allocators of PyTorch and onnxruntime and an allocator implemented by Greedy by Size for Offset Calculation (GSOC) algorithm proposed in work ~\cite{pisarchyk2020efficient}, which has achieved the near-optimal memory footprint for fixed-length input inference.
Memory footprint results are reported in Figure ~\ref{fig:intermediatefootprint} and the memory allocation results are reported in  Figure ~\ref{fig:allocateintermeddiat}, which are evaluated using C++ APIs.
The memory allocated for intermediate tensors in PyTorch and onnxruntime keep increasing during benchmarking.
After processing a long sequence request, 460 in this case, the memory usage reaches its peak and no longer drops.
The two runtimes incrementally build up a cache of CUDA memory and reassign it to later allocations.
When there are no available fragments in the cached memory blocks, they allocate a large block of additional memory and never release it until the memory usage reaches an upper limit.
In contrast, our allocator and GCOS use the information of computation-graph to control memory usage wisely.
As shown in the figure, the maximum memory usage of us is 12.15 MB.
The memory footprint of Turbo is quite close to the GCOS method.
However, Turbo allocates and frees less memory than GCOS for each inference.

We also analyze the overall memory footprint of the runtime.
The peak device memory usage is measured by monitoring the nvidia-smi information in every 1 ms.
the peak GPU memory used for Turbo is 663 MB while the GPU peak memory used by PyTorch is 1307 MB and 1653 MB for onnxruntime.
Since the CUDA contexts of a variety of CUDA kernels need to take up a large amount of memory space~\footnote{https://github.com/pytorch/pytorch/issues/20532}, PyTorch and onnxruntime require more GPU memory than the DNN model and intermediate tensors actually used, which valid our efforts to implement a lightweight transformer-target inference runtime instead of a training framework or a general inference runtime.

\begin{figure}[ht!]
\centering
\includegraphics[width=0.5\textwidth]{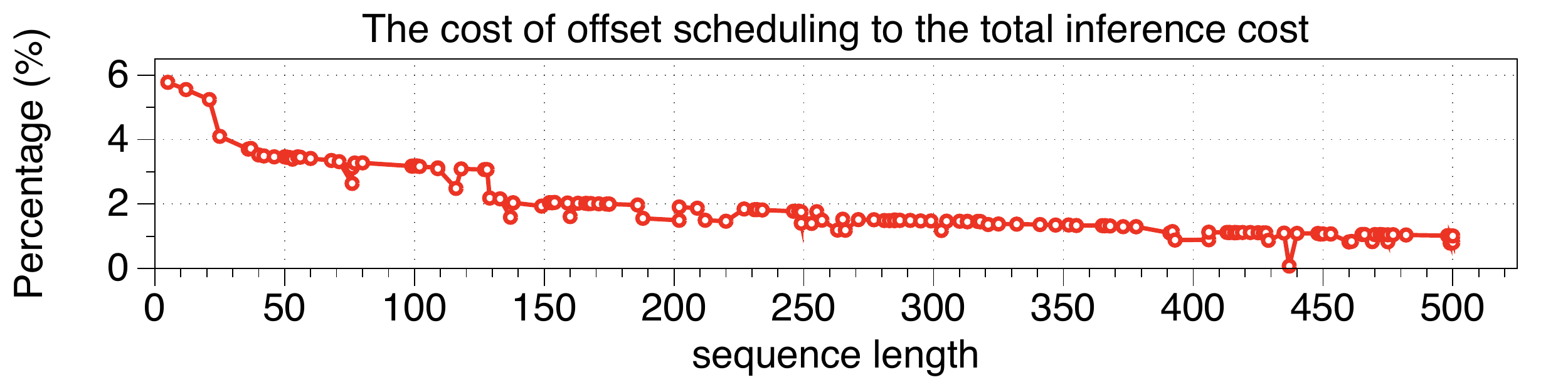}
\caption{The offset scheduling overhead of model-aware allocator.} 
\label{fig:allocator-overhead}
\end{figure}

To prove our proposed model-aware allocator's efficiency, we measure the overhead of Algorithm ~\ref{algo:allocator} to schedule memory offset of each intermediate tensor.
Since the time complexity of the algorithm is O(\#$tensor^2$),  a trick to reduce the number of tensors is that, for the DNN model with repeated structures, 
we only compute memory addresses once for intermediate tensors in one of those structures and reuse them in the other ones.
We measure the cost of Algorithm ~\ref{algo:allocator} in each inference process using a BERT model with input sequence lengths randomly generated from 5 to 500. 
As shown in Fig. ~\ref{fig:allocator-overhead}., the average cost of offset scheduling (cost of Algorithm ~\ref{algo:allocator}) is 1.8\% on average (0.07\%-5.77\%) of DNN inference latency. 
The overhead of the algorithm is extremely low and its benefit outweighs its cost.

\subsubsection{End-to-End Speed on Fixed-length Requests}
We also compared with another three popular runtimes that only support fixed-length requests.
1) TensorFlow-XLA is implemented with TensorFlow (version 1.13) and preprocessed with XLA.
2) Faster Transformers (v1) is a transformer boost software developed by NVIDIA, which implements a set of customized fused kernels like us.
However, it has no memory management and using the memory allocator of the TensorFlow framework.
3) NVIDIA TensorRT (v5.1.5) is an SDK for high-performance deep learning inference.
The above three solutions require a time-consuming pre-tuning process based on the input dimension in advance, so they cannot be applied to handling real-time variable-length requests.

\begin{figure}[ht!]
\includegraphics[width=0.5\textwidth]{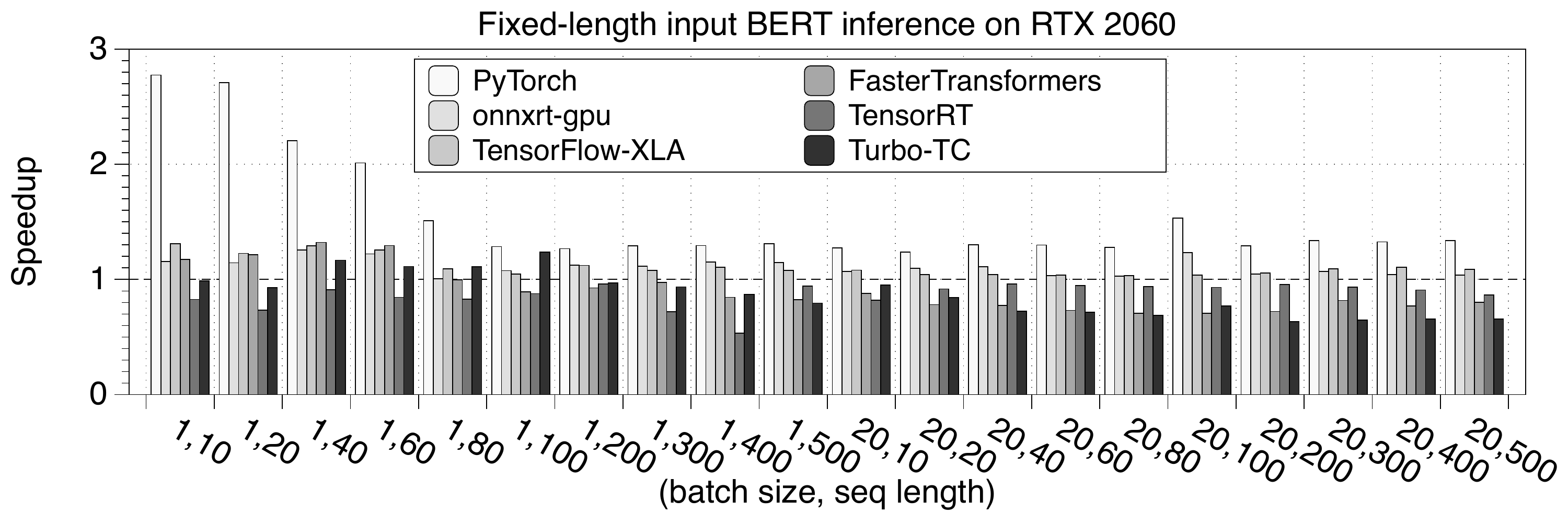}
\label{fig:2060-spd}
\caption{Benchmarking the runtimes for fixed-length input tasks on RTX 2060. The y-axis indicates normalized speedup of TurboTransformers.}
\label{fig:fixlength}
\end{figure}

For the sake of fairness, we chose BERT as the transformer model to be evaluated on a fixed-length input task, for which every runtime has been specifically optimized for it, and an official example is provided.
We select a parameter space consisting of the Cartesian Product of a set of batch sizes including 1 and 20 and a set of sequence lengths sampled from 10 to 500.
The speedups of TurboTransformers to the other runtimes are shown in Figure~\ref{fig:fixlength}.
Compared with PyTorch, the speedup of Turbo is 1.23x-2.77x, on average 1.54x.
Compared with onnruntime-gpu, the speedup of Turbo is 1.01x-1.26x, on average 1.11x.
Compared with TensorFlow-XLA, the speedup of Turbo is 1.03x-1.31x, on average 1.11x.
Compared with Faster Transformers, the speedup of Turbo is 0.71x-1.32x, on average 0.91x.
Compared with TensorRT, the speedup of Turbo is 0.53x-0.96x, on average 0.87x.
On average, our runtime is around 10\% faster than XLA and onnxruntime and around 10\% slower than Faster Transformers and TensorRT.
TensorRT needs an offline tuning process, during which it can select the optimal parameters for GEMM kernels and may identify the optimal CUDA thread block sizes for non-GEMM kernels.
On the contrary, our runtime does not involve the benefits of these optimizations.

\subsection{Performance of Serving Framework}
We choose a BERT-based service as our target application, which is used to classify a paragraph of text.
The input text is first randomly generated with sequence length uniformly distributed between 2-100 and then generated with sequence length uniformly distributed between 5-500. 
The requests are sent to the serving system with ~\textit{Poisson} inter-arrival times.
We turn off the caching optimization.

There are two strategies to build the cached\_cost dictionary in Algorithm~\ref{algo:batchscheduler}.
First, if the parameter space is small, a warmup phase that records the latency of executing with all possible parameters is required after the service started.
It takes tens of minutes in our experiments.
Second, if the parameter space is large, we sample the parameter space and use the interpolation method to estimate a specific case during serving. 
After you get real data, it can be used to update the dictionary in a lazy evaluation way. 

\begin{figure}[ht!]
\centering
\includegraphics[width=0.5\textwidth]{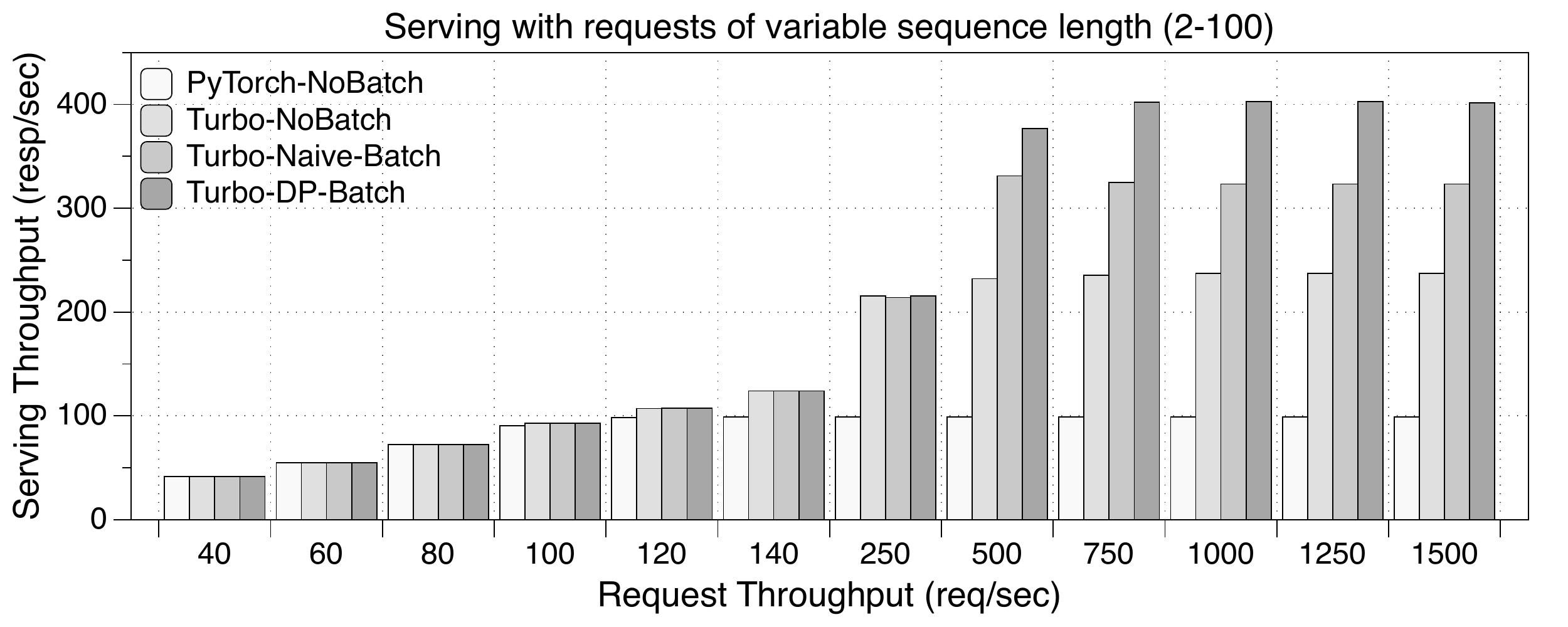}
\caption{Serving throughput under different request throughput (Sequence Length 2-100).}
\label{fig:servingthroughputsmall}
\end{figure}

The serving throughput of input sequence length ranging 2-100 is illustrated in Figure ~\ref{fig:servingthroughputsmall}.
The figure's x-axis indicates how many requests arrive at the serving system per second, ranging from 20 req/sec to 1500 req/sec.
The figure's y-axis represents how many responses can be obtained per second, which is usually called serving throughput.
The \textit{Critical Point} for service latency to remain stable is that request throughput, and serving throughput are equal.
When request throughput is higher than the critical point, the requests will accumulate in the message queue, leading to long delays of latter requests.
After a while, its latency will gradually tend to infinity ($+\infty$), and the service system has to drop some requests.

Our proposed batch scheduler achieves the best serving throughput.
The baseline is shown as PyTorch-Nobatch, which uses PyTorch as the runtime and serve without batching optimization.
Turbo-NoBatch is the service using our proposed runtime to replace PyTorch.
Turbo-Naive-Batch is implemented with a naive batch scheduler, which packs the requests currently inside the message queue into a single batch.
Turbo-DP-Batch is our proposed variable-length-aware batch scheduler.
The batch scheduler of our system employs the hungry strategy, and the maximum batch size is 20.
The serving throughput of PyTorch-Nobatch is saturated at 99 resp/sec.
The serving throughput of Turbo-NoBatch is saturated at 237 resp/sec (\textbf{2.39x}).
Turbo-Naive-Batch improves it to 323 resp/sec (\textbf{3.26x}) and the Turbo-DP-Batch futher improves it to 402 resp/sec (\textbf{4.06x}).

The latency results on four systems' critical points are shown in Table ~\ref{tab:latencysmall}.
Since batching brings higher GPU utilization for short requests, the latency of Naive-Batch is smaller than NoBatch.
Turbo-DP-Batch sorts the requests in MQ, the execution of long sequences is delayed, thus increasing their latency.
Therefore Naive-Batch brings smaller latency than Turbo-DP-Batch.

\begin{table}[ht!]
\tiny
\begin{centering}
\caption{The latency of four serving systems (sequence length 2-100). Table item contains the avg (min, max) latency in ms.}
\label{tab:latencysmall}
\begin{tabular}{l|ccccc}
\hline
Request & PyTorch & \multicolumn{3}{|c}{Turbo} \\
Thrpt (req/sec) & NoBatch & \multicolumn{1}{|c}{NoBatch} &  Naive-Batch & Turbo-DP-Batch  \\
\hline
\multirow{2}{*}{99} &  36.52 & 7.47   & {7.51}  & \textbf{7.49}  & \\
& (10.05, 108.72) & (3.84, \textbf{18.59}) & (\textbf{3.78}, {21.58}) & ({3.82}, 20.23) & \\
\hline
\multirow{2}{*}{237} & \multirow{2}{*}{+$\infty$}  & 16.68  & \textbf{10.09} & {10.89}  & \\
  & & (3.68, 45.39)  & (\textbf{3.67, 28.60}) & {(3.81, 37.00)}  & \\
\hline
\multirow{2}{*}{323}  & \multirow{2}{*}{+$\infty$} & \multirow{2}{*}{+$\infty$} & \textbf{12.44} & {15.37} & \\
& & & \textbf{(3.48, 26.70)} & {(3.58, 40.45)} & \\
\hline
\multirow{2}{*}{402} & \multirow{2}{*}{+$\infty$} & \multirow{2}{*}{+$\infty$} & \multirow{2}{*}{+$\infty$} & \textbf{24.74} & \\
 &  &  & & \textbf{(4.23, 57.02)} & \\
\hline
\end{tabular}
\end{centering}
\end{table}

\begin{figure}[ht!]
\centering
\includegraphics[width=0.5\textwidth]{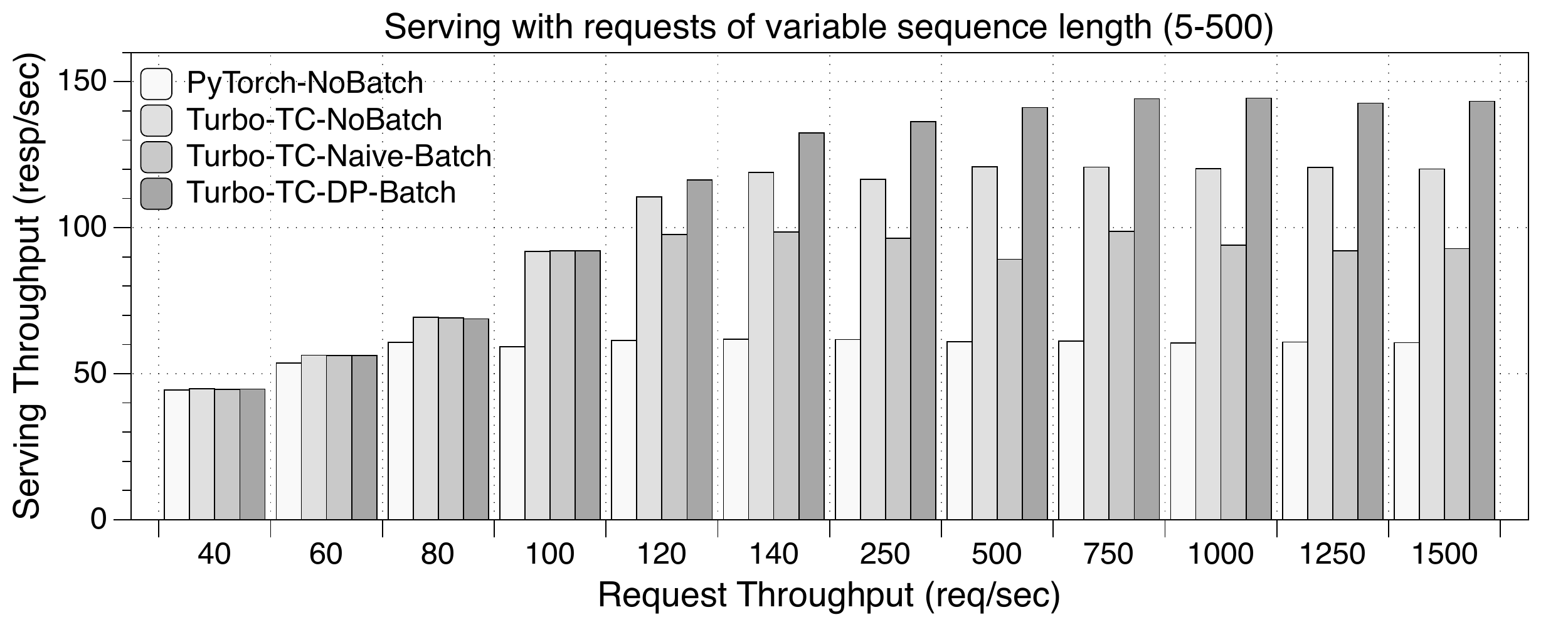}
\caption{Serving throughput under different request throughput (Sequence Length 5-500).}
\label{fig:servingthroughputlarge}
\end{figure}

When increasing the length difference between the sequences by changing the input sequence range from 2-100 to 5-500, our proposed allocator achieves the best throughput and leads to the lowest latency.
The throughput of input sequence length ranging 5-500 is illustrated in Figure ~\ref{fig:servingthroughputlarge} and the latency results are shown in Table~\ref{tab:latencylarge}.
In this case, we turn on the tensor core optimization, which brings no accuracy loss and can better satisfy the SLO.
The serving throughput of PyTorch-Nobatch is saturated at 60 resp/sec, while Turbo-TC-NoBatch improves it to 120 resp/sec (\textbf{2.0x}).
Turbo-TC-DP-Batch further improves it to 144 resp/sec (\textbf{2.4x}).
Due to the additional zero-padding overhead, Turbo-TC-Naive-Batch's throughput is 98 resp/sec, which is even worse than Turbo-NoBatch.
As shown in Table~\ref{tab:latencylarge}. Under the same request throughput rate, Turbo-DP-Batch usually brings the lowest average and maximum service latency.
This is because of the reduction of zero-padding cost and shortening the latency of every single request.


\begin{table}[ht!]
\tiny
\begin{centering}
\caption{The latency of four serving systems (sequence length 5-500). Table item contains the avg (min, max) latency in ms.}
\label{tab:latencylarge}
\begin{tabular}{l|ccccc}
\hline
Request & PyTorch & \multicolumn{3}{|c}{Turbo} \\
Thrpt (req/sec) & NoBatch & \multicolumn{1}{|c}{NaiveBatch} & NoBatch & Turbo-DP-Batch  \\
\hline
\multirow{2}{*}{60} &  77.71 &17.80   & \textbf{8.05}  & \textbf{8.05}  & \\
& (10.61, 158.06) & (3.06, 121.96) & (2.76, \textbf{20.53}) & (\textbf{2.70}, 27.42) & \\
\hline
\multirow{2}{*}{98} & \multirow{2}{*}{+$\infty$}  & 16.68  & 24.88 & \textbf{13.79}  & \\
  & & (2.96, 65.09)  &  (3.0, 65.09) & \textbf{ (2.94, 45.09)}  & \\
\hline
\multirow{2}{*}{120}  & \multirow{2}{*}{+$\infty$} & \multirow{2}{*}{+$\infty$} & 32.91 & \textbf{23.18} & \\
& & & (3.14, 127.61) & \textbf{(2.72, 81.83)} & \\
\hline
\multirow{2}{*}{144} & \multirow{2}{*}{+$\infty$} & \multirow{2}{*}{+$\infty$} & \multirow{2}{*}{+$\infty$} & \textbf{38.51} & \\
 &  &  & & \textbf{(4.44, 106.65)} & \\
\hline
\end{tabular}
\end{centering}
\end{table}

\section{Conclusion}
TurboTransformers improves latency and throughput for deploying transformer-based DL services in GPU datacenter.
It solves two critical problems introduced by the transformer models, which are unprecedented computation pressure and variable-length input.
For these purposes, it proposed three innovations in computing, memory and serving levels, 
including a new parallel batch reduction algorithm for Softmax and LayerNorm kernels,
a sequence-length-aware memory allocator as well as a sequence-length-aware batch scheduler.
The runtime achieves better speed than PyTorch and similar speed as onnxruntime in the variable-length request tests but maintains a smaller memory footprint.
It also achieves comparable speed than TensorFlow-XLA, TensorRT, and FasterTransformers in the fixed-length request tests.
While conventional batching is inefficient for variable-length request, the serving framework achieves higher throughput using the proposed batch scheduler.

\section{Acknowledgements}
We would like to thanks Yin Li (UC Davis), Shengqi Chen, Wentao Han (Tsinghua Univ.) for their proofreading.

\bibliography{my.bib}

\appendix
\section{Artifact Appendix}
\subsection{Abstract}
The artifact contains the code for the runtime of TurboTransformers. 
We provide instructions for obtaining critical results used in this paper and scripts for running the experiments in the paper.

\subsection{Description}
\subsubsection{Check-list (artifact meta information)}
\begin{itemize}
\item
\textbf{Algorithms}: The artifact includes the runtime component of TurboTransformers.
More specifically, it contains the parallel batch-reduction algorithms proposed in Section ~\ref{sec:gpu_batch_reduction}, as well as the model-aware-allocator proposed in Section~\ref{sec:memory_manager}.
\item 
\textbf{Datasets}: 
The inputs used for benchmark scripts of the artifact are randomly generated.
\item 
\textbf{Compilation}: 
The experiments in the paper used g++ version 7.5.0, nvcc version 10.2.
The artifact also provided a docker file to build a docker environment to compile the code.
The docker version used is 19.03.8.
\item
\textbf{Runtime environment}:
The artifact provided a compilation script to run on the CentOS 7 OS installed with CUDA version 10.2.
\item
\textbf{Hardware}: 
The artifact can run on a server equipped with at least one NVIDIA GPU.
The supported  GPU microarchitecture codenames include Pascal, Volta, Turing.
\item
\textbf{Output}: Running times of the algorithms are output
to the console.
\item
\textbf{Experiment Workflow}:
Clone the repository and use the provided scripts to run the experiments.
\item
\textbf{Publicly available}: Yes
\end{itemize}

\subsubsection{How Delivered}
Available as open-source under the BSD license: https://github.com/Tencent/TurboTransformers.
The artifact branch is ppopp21\_artifact\_centos.

\subsection{Installation}
\subsubsection{Docker}
\begin{itemize}
\item
Build a docker image using the docker file.

\begin{lstlisting}[language=bash]
bash tools/build_docker_gpu.sh $PWD
\end{lstlisting}
\item
Run the image as a container

\begin{lstlisting}[language=bash]
nvidia-docker run --gpus all --net=host --rm -it -v $PWD:/workspace --name=turbo_dev:latest
\end{lstlisting}

\item
Inside the container, build the artifact.

\begin{lstlisting}[language=bash]
cd /workspace
bash tools/build_and_run_unittests.sh $PWD -DWITH_GPU=ON
\end{lstlisting}

\end{itemize}

\subsubsection{CentOS}
\begin{itemize}
\item
Use the following command to build the artifact on CenOS 7.
\begin{lstlisting}[language=bash]
bash build_centos.sh
\end{lstlisting}
\end{itemize}

\subsection{Experiment Workflow}
You can run the scripts provided in the benchmark directory to compare the speed of turbo runtime with PyTorch.
Run the following script will reproduce the Bert and Albert results shown in Figure ~\ref{fig:variable-latency}.

\begin{lstlisting}[language=bash]
bash gpu_run_variable_benchmark.sh
\end{lstlisting}

Run the following script will reproduce the Bert and Albert result in Figure ~\ref{fig:fixlength}.

\begin{lstlisting}[language=bash]
bash gpu_run_fixed_benchmark.sh
\end{lstlisting}

\end{document}